\documentclass[aps,amsmath,showpacs,amsfonts,10pt]{revtex4}
\usepackage{epsfig,graphicx}
\usepackage[english]{babel}
\usepackage{amsfonts}
\usepackage{amsmath}
\usepackage{latexsym}
\usepackage{graphics,bm}
\usepackage{natbib}
\usepackage{dcolumn}
\usepackage{bm}
\usepackage{rotating}

\begin{document}

\title{Dynamics of an optomechanical resonator containing a Quantum Well induced by periodic modulation of cavity field and external laser beam}

\author{ Sonam Mahajan$^{1}$ , Neha Aggarwal$^{1,2}$, Tarun Kumar$^{3}$, Aranya B Bhattacherjee$^{2,4}$ and ManMohan$^{1}$}

\address{$^{1}$Department of Physics and Astrophysics, University of Delhi, Delhi-110007, India} \address{$^{2}$Department of Physics, ARSD College, University of Delhi (South Campus), New Delhi-110021, India}
\address{$^{3}$Department of Physics, Ramjas College, University of Delhi , Delhi-110007, India}
\address{$^{4}$School of Physical Sciences, Jawaharlal Nehru University, New Delhi-110067, India}

\begin{abstract}
We study in detail the dynamics of a non-stationary system composed of a Quantum Well confined in an optomechanical cavity. This system is investigated with classical and quantized motion of the cavity movable mirror. In both the cases, the cavity frequency is rapidly modulated in time. The resultant periodically modulated spectra is presented. In particular, we study the effect of two-photon process on the number of intracavity photons. The intensity of fluorescent light emitted by excitons in the quantum well is also examined for these non-stationary systems. It is observed that the initial stage of fluorescence spectrum helps in detecting the two-photon process. It is also noticed that under strong modulation, two-photon process dominates while under weak modulation, fluorescence dominates. We also analyzed the dynamics of the system induced by a periodic modulation of the external pump laser with constant cavity frequency. This modulation of external laser pump helps in phonon amplification.
\end{abstract}


\maketitle

\section{Introduction}
Study of matter-light interaction in semiconductor nano-structures like quantum dots and quantum well (QW) have led to a large variety of interesting phenomena like the Autler-Townes doublet \citep{aut} and vacuum Rabi splitting \citep{agg, tho, khi}. The quantum properties of light like antibunching, squeezing and bistability are also exhibited by quantum systems due to the synonymity between excitonic and atomic resonances \citep{kim, man, baa, mes}. The cavity quantum electrodynamics (QED) in semiconductor systems like quantum wells and quantum dots have potential application in opto-electronic devices \citep{shi, ele1, ele2}. The cavity QED in atomic systems exhibit antibunching whereas bunching is observed in the fluorescent spectrum of the QW \citep{ere, vya}. 

The cavity quantum optomechanics is the latest field of research which has grabbed attention in a vast variety of systems ranging from gravitational wave detectors (LIGO project) \citep{cor, cor1}, nanomechanical cantilevers \citep{hoh, gig, arc, kle, fav, reg}, vibrating microtoroids \citep{carm, sch1}, membranes \citep{jdt} and ultracold atoms \citep{bre, mur, bha, bha1, tre}. In recent years, there has been a continuous growing interest in cooling of optomechanical systems to their quantum ground state \citep{vit1, vit2, vit3, tar2, son1, son2}. This field has a potential application in a large variety of sensitive measurements like detection of weak forces \citep{bra1, abr}, small masses \citep{jen} and small displacements \citep{lat}. In a pioneering work \citep{bra}, it has been predicted that the dynamics of back-action arises due to the radiation pressure force by the optical field on the moving mirror of optomechanical systems. The radiation pressure exerted by an optical field on the moving end mirror of the cavity gives rise to the coupling between the intensity of the intracavity light field and displacement of the mirror. This nonlinearity arises in the system due to change in the path of light in an intensity-dependent way. Such nonlinear effects have been studied for an optomechanical system consisting of a QW \citep{eyo}. The non-linearity analogous to Kerr nonlinearity can be achieved in an optomechanical system \citep{tar1}. 

Two decades ago, Moore \citep{moo} discussed the quantization of electromagnetic field in a cavity with movable perfectly reflecting boundaries. Later, Dodonov and co-workers \citep{dodc, dodc1} generalized the Moore's theory so that the effects of time-varying refractive index of the medium inside the cavity were also included. The main interest in such kind of system is the creation of photons due to the nonadiabatic distortion of the electromagnetic vacuum state. It has been presumed that a moving boundary with nonuniform motion \citep{dew, ful} or a fast changes in the refractive index of the cavity medium \citep{yab, hiz} can generate real photons from the vacuum state. This phenomenon is nowadays termed as dynamical Casimir effect which was apparently introduced by Yablonovitch \citep{yab} and Schwinger \citep{sch}. Theoretically, an effective Hamiltonian was derived for this type of system consisting of radiation inside a cavity with moving boundary and a time-varying dielectric medium in the cavity \citep{law}. It was predicted that a nonadiabatic process characterized by two-photon process helps in creating photon pairs from the vacuum state. It was identified that this two-photon character of the optical field is related to the squeezing phenomenon \citep{sar, dod1}. Further, parametric process can be observed in an optomechanical system by periodically modulating the external driving field which is also efficient in squeezing phenomenon \citep{law2, mar2}. 

Motivated by these interesting features in the field of non-stationary cavity QED, we propose a system containing a quantum well within an optomechanical cavity undergoing the periodic oscillations. We investigate this non-stationary system with classical mirror motion and quantized mirror motion. The cavity frequency is periodically modulated with time. We examined the fluorescent intensity of light emitted by exciton in QW and the average number of photons within the cavity. In the end, we have proposed a quantum system composed of a QW inside an optomechanical cavity with the fixed cavity frequency and the amplitude modulated external laser pump. In this case, we observe the amplification of phonons inside the cavity.         
   
\section{Optomechanical Cavity with Classical Mirror Motion}

\begin{figure}[h]
\hspace{-0.0cm}
\includegraphics [scale=0.8]{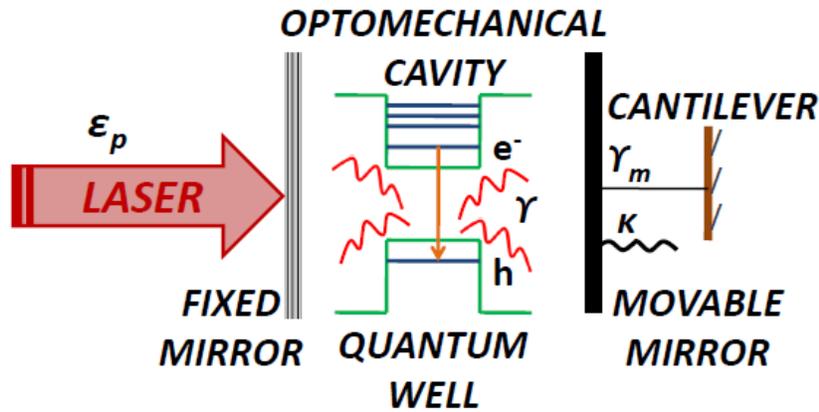}
\caption{(color online) Schematic figure of the system consisting of Quantum Well in an Optomechanical Cavity with one fixed mirror and other movable mirror attached to Cantilever.}
\label{b}
\end{figure}

In this section, we introduce the basic model of the system with QW in an optomechanical cavity. The cavity has one mirror fixed and other mirror movable as shown in figure \ref{b}. The system investigated here is composed of a semiconductor QW in an optomechanical cavity driven by an external pump laser and interacting with a single-mode optical field. The optomechanical cavity has frequency $\omega_{c}$ which is driven by an external coherent light field with frequency  $\omega_{p}$. 
The exciton-exciton scattering is ignored as the density of excitons is small which restricts the exciton-cavity mode to a linear regime. In this section, we treat the motion of the movable mirror as a classical harmonic oscillator. Also, the dynamics of this system is studied with time varying cavity frequency. In the present system, the one-dimensional single quantized cavity mode has time-dependent frequency $\omega_{c}\left(t\right)=\omega_{c}\left(1+\epsilon\sin\left(\Omega t\right)\right)$ with unperturbed frequency $\omega_{c}$. Here $\epsilon$ is the modulation amplitude and $\Omega$ is the frequency of the periodic modulation used for time varying cavity frequency. This particular form of time dependent cavity frequency is a result of the harmonic motion of the cavity mirror. The time-dependent Hamiltonian of the system in the rotating-wave and dipole approximation is given by \citep{eyo, law}

\begin{eqnarray}\label{one}
H_{I}=\hbar\omega_{b}b^{\dagger}b+\hbar\omega_{c}(t)a^{\dagger}a+\hbar g_{0}\left(a^{\dagger}b+b^{\dagger}a\right)+i\hbar\chi\left(t\right) \left(a^{\dagger 2}e^{-2i\omega_{p}t}-a^{2}e^{2i\omega_{p}t}\right)+i\hbar\epsilon_{p}\left(a^{\dagger}e^{-i\omega_{p}t}-a e^{i\omega_{p}t}\right).  
\end{eqnarray}

The first term describes the free energy of exciton in the QW. Here $b(b^{\dagger})$ is the exciton annihilation (creation) operator obeying the commutation relation $\left(\left[b,b^{\dagger}\right]=1\right)$ and $\omega_{b}$ is the exciton frequency in the QW. The second term is the energy of the single optical mode with $a(a^{\dagger})$ being the lowering (raising) operator for the photon in the cavity which follows the commutation relation as $\left(\left[a,a^{\dagger}\right]=1\right)$. The third term depicts the interaction energy between the exciton and photon. Here $g_{0}$ is the coupling constant for the exciton-photon interaction energy. The fourth term is due to the two-photon process that arises because of the time-varying cavity field. This term has close connection with the Hamiltonian of the harmonic oscillator with a time-dependent frequency \citep{law}. Here $\chi(t)$ is the effective frequency which is arbitrary function of time. The last term in the Hamiltonian is the energy due to the external pump laser where $\epsilon_{p}$ is the amplitude of laser pump.

Now we will discuss in brief about the fourth term in the Hamiltonian. There are normally two types of nonadiabatic processes in a system composed of an optical cavity with one mirror fixed and other mirror movable as mentioned in ref. \citep{law}. First one is characterized by $a^{\dagger}a$ terms in the Hamiltonian. In this type of process, the total number of photons inside the cavity remains same. The second one is characterized by the terms $a^{\dagger 2}$ or ${a^{2}}$. In this process, photon pairs can be created from the vacuum state. This kind of process is known as two-photon process. This two-photon process is responsible for the fourth term in the Hamiltonian $\left(\ref{one}\right)$. It is related to the squeezing phenomenon \citep{plu}. Due to the fact that the two-photon character of the optical field helps in creating the photon pairs from the vacuum state, therefore, it is also used to study the Dynamical Casimir Effect in various systems \citep{dod5, dod2, dod3}. 

The dynamics of the system is fully described when the fluctuation dissipation processes affecting the optical and excitonic modes are included into the system. The interaction of the system with external degrees of freedom brings the dissipation into the system. The leakage of photons through the mirrors damps the optomechanical cavity field. $\kappa$ is the decay constant of cavity light field.The spontaneous decay rate of exciton is $\gamma$. The full dynamics of the system can be described by the following set of quantum Langevin equations (QLE) by taking into account all the dissipative processes in the frame rotating at the frequency $\omega_{p}$:

\begin{equation}\label{two}
\dot{b}=-i\delta_{b}b-ig_{0}a-\gamma b+\sqrt{2\gamma}b_{in},
\end{equation}   

\begin{equation}\label{three}
\dot{a}=-i\Delta a-i\omega_{c}\epsilon\sin\left(\Omega t \right)a-ig_{0}b+2\chi(t)a^{\dagger}+\epsilon_{p}-\kappa a+\sqrt{2\kappa}a_{in},
\end{equation} 

where $\delta_{b}=\omega_{b}-\omega_{p}$ is the exciton-pump detuning and $\Delta=\omega_{c}-\omega_{p}$ is the cavity-pump detuning. In the above equations, $b_{in}$ and $a_{in}$ represent the input vacuum noise of exciton and cavity photons respectively. Their non zero correlation functions are given as follows \citep{eyo, eyo21}

\begin{equation}\label{four}
\left\langle b_{in}(t)b_{in}^{\dagger}(t')\right\rangle =\left\langle a_{in}(t)a_{in}^{\dagger}(t')\right\rangle =\delta\left(t-t'\right). 
\end{equation}

In various studies, there is a relation between the functions $\omega_{c}(t)$ and $\chi(t)$ which is given as \citep{law}

\begin{equation}\label{fivea}
\chi(t)=\frac{1}{4\omega_{c}(t)}\frac{d\omega_{c}(t)}{dt}.
\end{equation}

The time modulation with small amplitude is considered for the realistic case i.e. we take $\vert\epsilon\vert\ll1$. Therefore from Eqn. \ref{fivea}, we obtain

\begin{equation}\label{fiveb}
\chi(t)\approx\frac{\epsilon \Omega}{4}\cos\left(\Omega t\right)\approx2\chi_{0}\cos\left(\Omega t \right),  
\end{equation}

where $\chi_{0}=\left(\epsilon\Omega\right)/8$. We solve the coupled QLE [Eqs. \ref{two} and \ref{three}] numerically using MATHEMATICA 9.0 to study the fluorescent intensity of light emitted by the excitons in the QW. The dynamics of the system is also studied via the evolution of power $\left(\left\langle a^{\dagger}a\right\rangle\right)$. We have investigated the system under the strong modulation $\left(\chi_{0}\gtrsim g_{0}\right)$ and weak modulation $\left(\chi_{0}\ll g_{0}\right)$ and in two different limits i.e. in the bad cavity limit $(\kappa\gg\gamma)$ in the regime $\kappa>\epsilon$ and in the good cavity limit $(\kappa\ll\gamma)$ in the regime $\kappa\lesssim\epsilon$. 

\begin{figure}[h]
\hspace{-0.0cm}
\begin{tabular}{cc}
\includegraphics [scale=0.75]{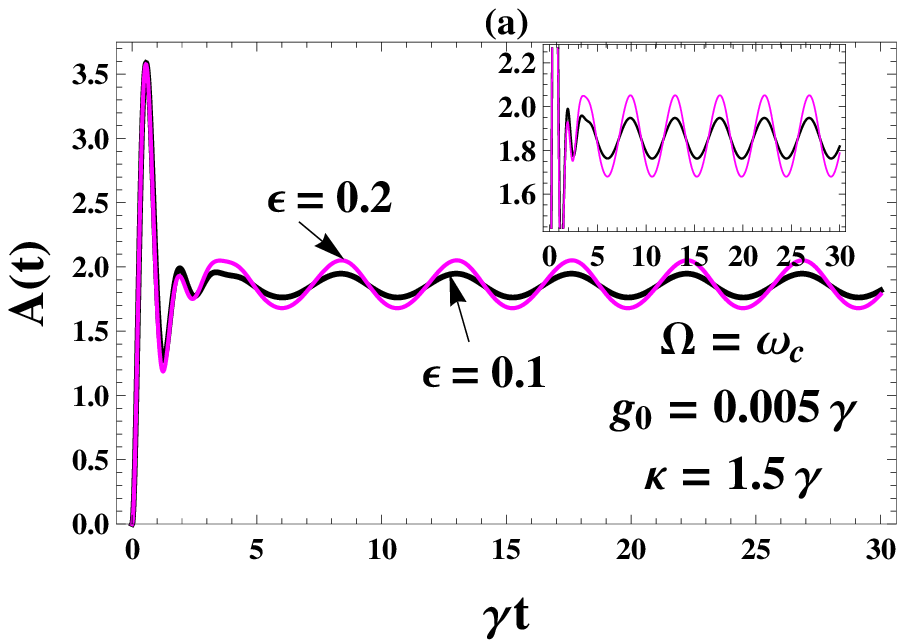}& \includegraphics [scale=0.75] {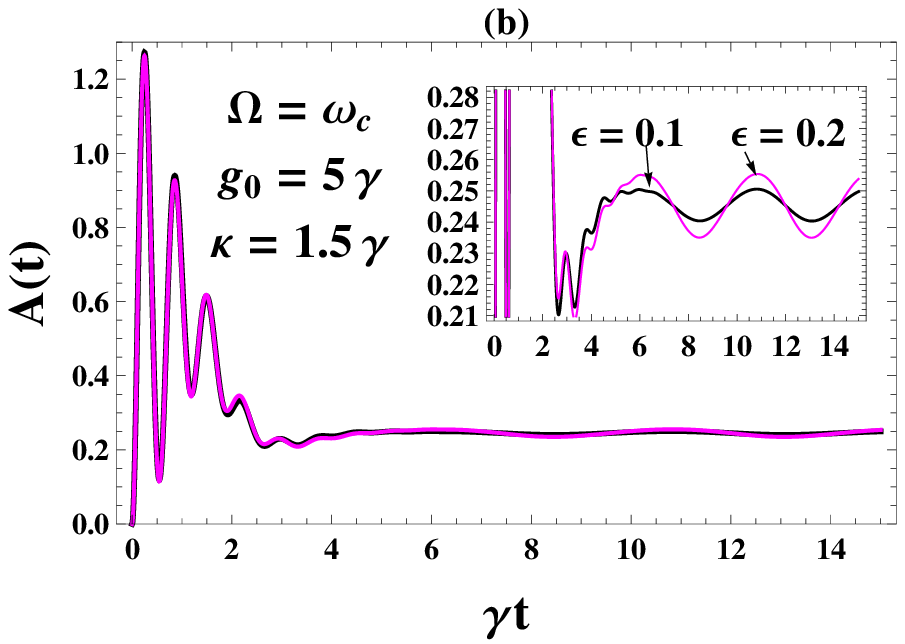}\\
\includegraphics [scale=0.75]{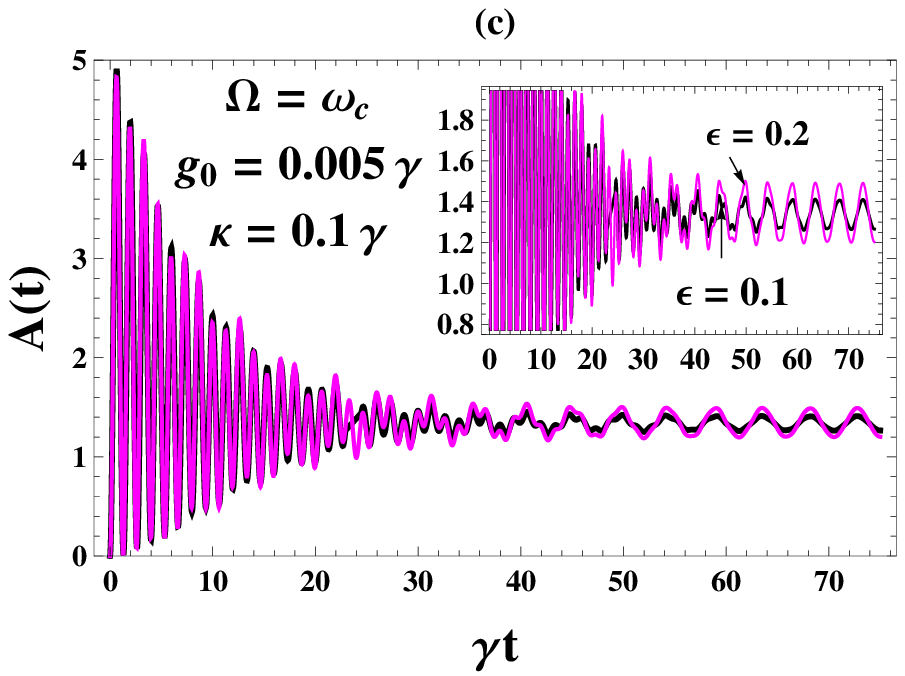}& \includegraphics [scale=0.75] {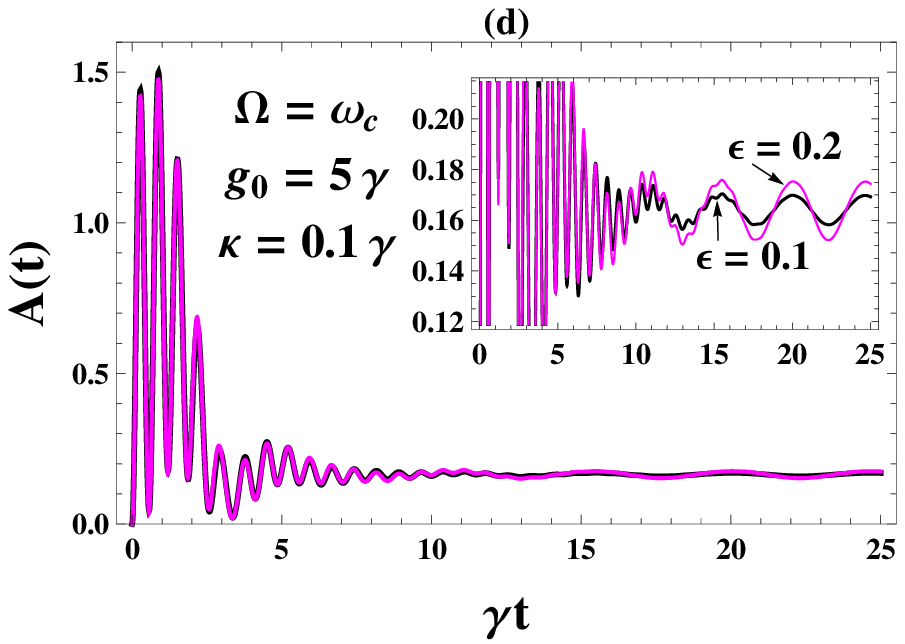}\\
 \end{tabular}
\caption{(Color online) Plots of intracavity mean number of photons $\left(A(t)=\langle a^{\dagger}a\rangle\right)$ inside an optomechanical cavity with classical mirror motion versus scaled time $\left(\gamma t\right)$ using time modulated cavity frequency at resonant frequency $\left(\Omega = \omega_{c}\right)$ for two modulation amplitudes $\epsilon = 0.1$ (thick line) and $\epsilon = 0.2$ (thin line). The parameters used are $\delta_{b} = 2\gamma$, $\Delta = 4.712\gamma$, $\epsilon_{p} = 5\gamma$, $\gamma_{m} = 10^{-5}\gamma$, $n_{th} = 175$ and $\omega_{c} =1.36\gamma$ . Plots (a) and (b) show the variation of $A(t)$ with time under the strong modulation $\left(g_{0} = 0.005\gamma\right)$ and weak modulation $\left(g_{0} = 5\gamma\right)$ respectively in the bad cavity limit for $\kappa>\epsilon \left(\kappa = 1.5\gamma\right)$. Plots (c) and (d) show the variation of $A(t)$ with time under the strong modulation $\left(g_{0} = 0.005\gamma\right)$ and weak modulation $\left(g_{0} = 5\gamma\right)$ in the good cavity limit for $\kappa\lesssim\epsilon \left(\kappa = 0.1\gamma\right)$.} 
\label{c}
\end{figure}

In figure \ref{c}, the intracavity photon number $\left(A(t)=\langle a^{\dagger}a\rangle\right)$ is plotted with scaled time $\left(\gamma t\right)$ under the strong and weak modulation regimes for both $\kappa>\epsilon$ and $\kappa\lesssim\epsilon$ at resonant modulating frequency $\left(\Omega=\omega_{c}\right)$ for two different modulation amplitudes $\epsilon = 0.1$ (thick line) and $\epsilon = 0.2$ (thin line). It is clearly seen from the plot that under the strong modulation, there is slight difference between the photon number for the two modulation amplitudes. For the case of weak modulation, there is not much deviation between the two plots. Under the strong modulation, periodic oscillations are observed with time. It shows an increase in the photon number inside the cavity due to the two-photon process. But for the weak modulation, the oscillations decay with time. This shows that the two-photon process is not dominant under the weak modulation. Initially, an increase in the intracavity photon number is noticed both for the weak and strong modulation. This is because initially excitons are in the excited state. They deexcite to the ground state by releasing photons in the cavity. In the case of strong modulation, the amplitude of initial peak is high as here photon number is amplified inside the cavity due to the two-photon process. Also, under the strong and weak modulations, we observe that increase in modulation amplitude enhances the average number of photons inside the cavity. Hence for a non-stationary system composed of a QW confined in an optomechanical cavity with classical mirror motion shows the periodic increase and decrease in photon number inside the cavity under the strong modulation but for the weak modulation, the photon number exhibits non-periodic damped oscillations. 

\begin{figure}[h]
\hspace{-0.0cm}
\begin{tabular}{cc}
\includegraphics [scale=0.75]{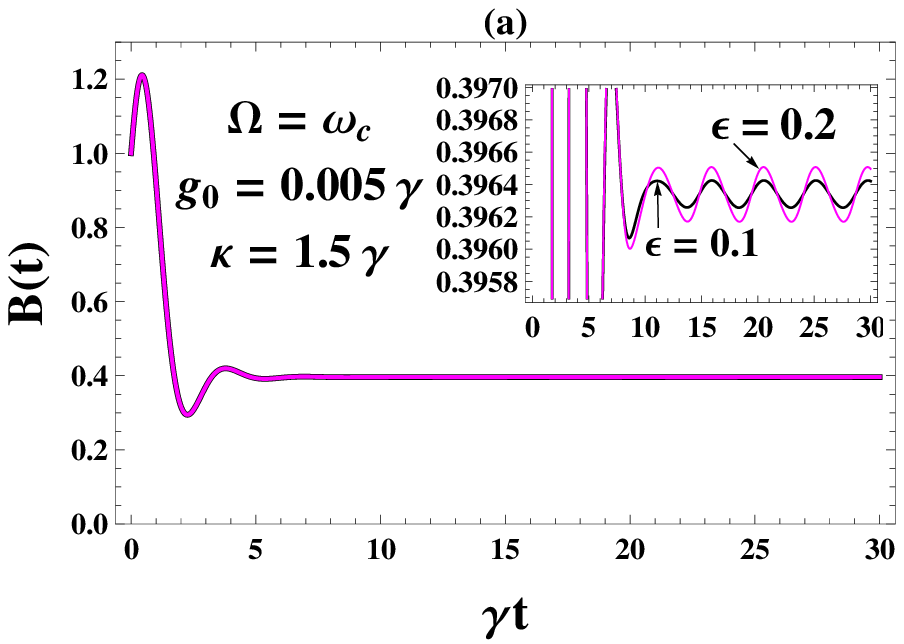}& \includegraphics [scale=0.75] {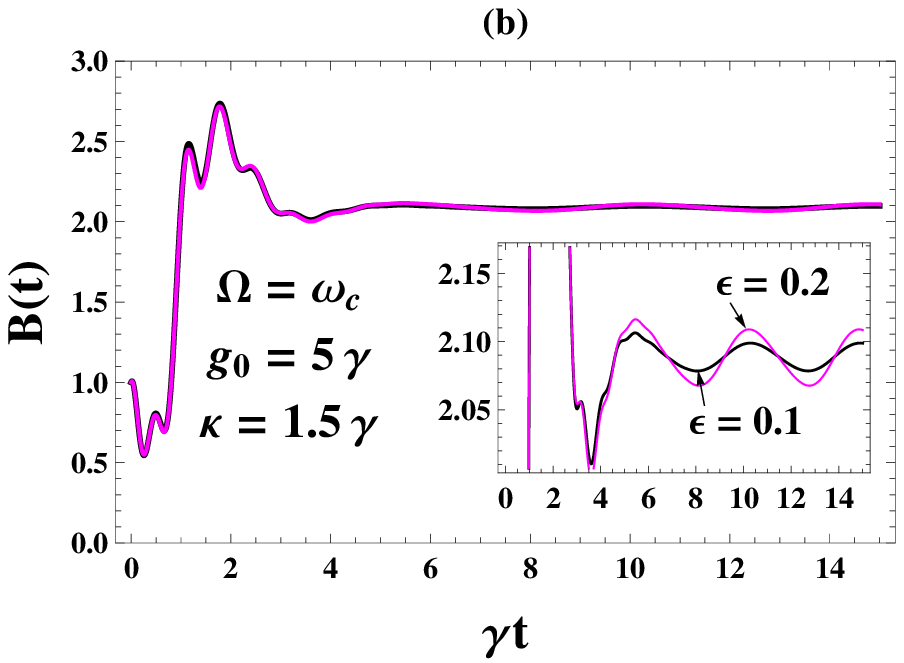}\\
\includegraphics [scale=0.75]{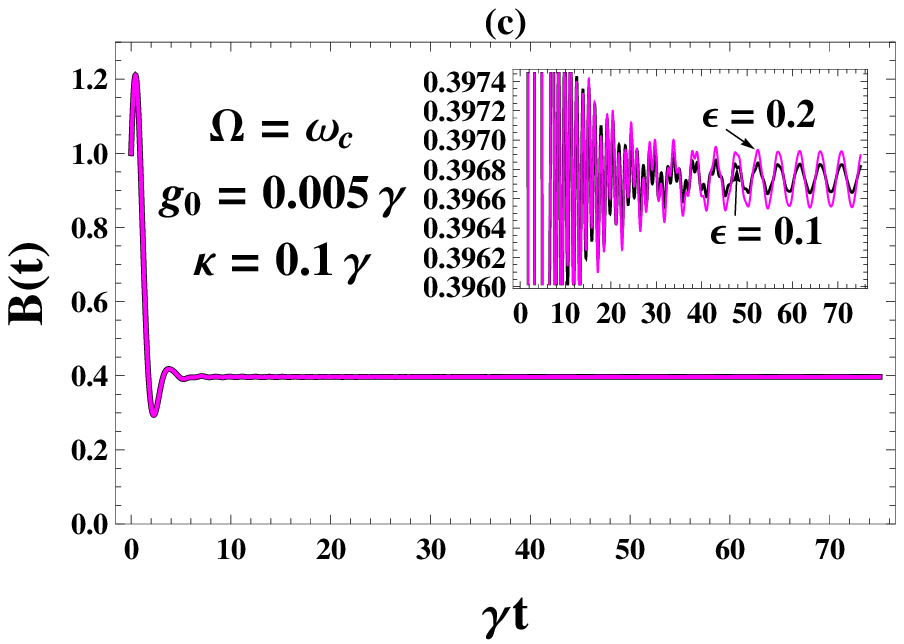}& \includegraphics [scale=0.75] {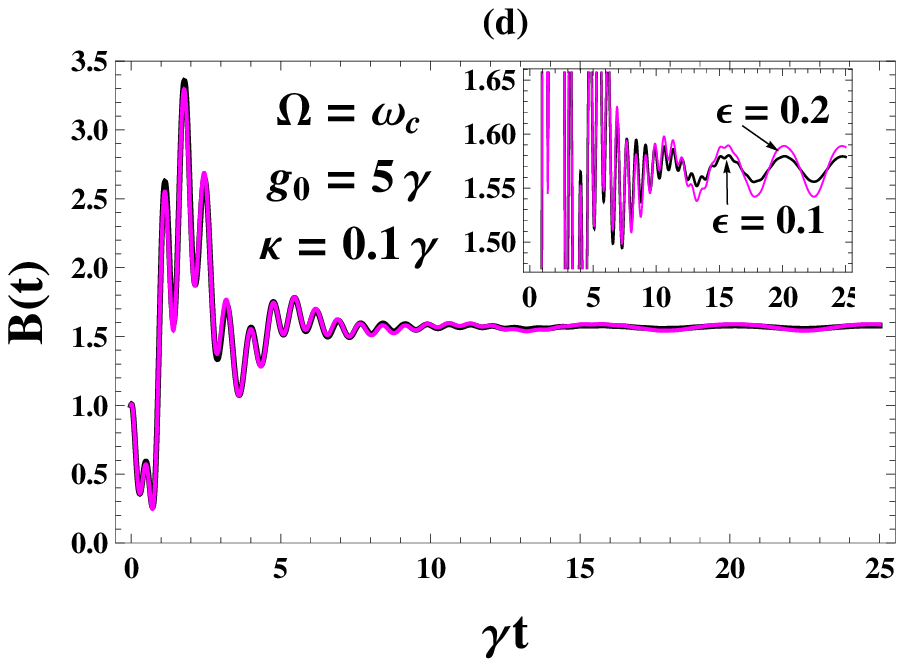}\\
 \end{tabular}
\caption{(Color online) Plots of intensity of fluorescent light $\left(B(t)=\langle b^{\dagger}b\rangle\right)$ versus scaled time $\left(\gamma t\right)$ inside an optomechanical cavity using time modulated cavity frequency at resonant frequency $\left(\Omega = \omega_{c}\right)$ for two modulation amplitudes $\epsilon = 0.1$ (thick line) and $\epsilon = 0.2$ (thin line). Plots (a) and (b) show the variation of $B(t)$ with time under the strong modulation $\left(g_{0} = 0.005\gamma\right)$ and weak modulation $\left(g_{0} = 5\gamma\right)$ respectively in the bad cavity limit for $\kappa>\epsilon \left(\kappa = 1.5\gamma\right)$. Plots (c) and (d) show the variation of $B(t)$ with time under the strong modulation $\left(g_{0} = 0.005\gamma\right)$ and weak modulation $\left(g_{0} = 5\gamma\right)$ in the good cavity limit for $\kappa\lesssim\epsilon \left(\kappa = 0.1\gamma\right)$. The other parameters used are same as in figure \ref{c}}.
\label{d}
\end{figure}

Figure \ref{d} shows the fluorescent spectrum of light $\left(B(t)=\langle b^{\dagger}b \rangle\right)$ with scaled time $\left(\gamma t\right)$ at resonant modulating frequency $\left(\Omega=\omega_{c}\right)$ under the strong and weak modulation regimes for both $\kappa>\epsilon$ and $\kappa\lesssim\epsilon$ for two different modulation amplitudes $\epsilon = 0.1$ (thick line) and $\epsilon = 0.2$ (thin line). The fluorescent light amplitude is proportional to the number of excitons. The figure depicts that the intensity of the fluorescent light increases as the modulation amplitude of time-modulated cavity frequency increases. However, the difference between the intensities is very small. Also, one can see that the intensity of fluorescent light decays with time. In general, the fluorescent spectrum exhibits nonperiodic damped oscillations. Moreover, there is an amplification in the mean number of excitons in the QW for the initial moment under strong modulation which gradually decreases and becomes stationary with time. However, under the weak modulation, the average number of excitons decreases initially, but gradually, the intracavity photons excite one or more excitons in the quantum well, leading to enhanced emission of fluorescence under the weak modulation. Furthermore, a saturation is observed in excitation of excitons for both strong and weak modulations. A system composed of a QW in an optical cavity with constant cavity frequency was considered \citep{eyo3}. It has been observed that the intensity of the fluorescent light emitted by excitons in the QW exhibits nonperiodic damped oscillations. Here, similar behaviour is observed in the oscillations of the fluorescent spectrum. 

We have also analyzed the effect on the mean number of intracavity photons $\left(A(t)\right)$ and fluorescent spectrum of light $\left(B(t)\right)$ by turning off the external driving field $\left( \epsilon_{p}=0\right)$. It was noticed that the behaviour of the intracavity photons $\left(A(t)\right)$ and fluorescent spectrum of light $\left(B(t)\right)$ remain same but their magnitude decreases. It should be noted here that fluorescence photons is directly proportional to number of excitons in the cavity. Comparing figures of fluorescent spectrum of light $\left(B(t)\right)$ under strong modulation (see figs. \ref{d}(a) and (c)) and weak modulation (see figs. \ref{d}(b) and (d)), one can notice that initially there is a rise in the mean number of excitons under the strong modulation and a corresponding dip under weak modulation. This clearly shows that under the strong modulation the character of two-photon dominates which amplifies photons in the cavity leading to enhancement in the number of excitons initially. However, under the weak modulation, the effect of two-photon process is less as a result of which there is a decrease in the initial number of excitons in the cavity. Similar result was observed in a system with constant cavity frequency \citep{eyo3}. Therefore, by looking at the fluorescent spectrum of light at initial stage, one could easily detect the two-photon process. As time progresses, it is also observed that the steady state value of intracavity photon number $\left(A(t)\right)$ due to the two-photon process is more than the photon number due to the fluorescence $\left(B(t)\right)$ under the strong modulation. However, under the weak modulation, the photon number due to two-photon process $\left(A(t)\right)$ inside the cavity is less as compared to the intracavity photon number due to the process of fluorescence $\left(B(t)\right)$. This can be explained as under the strong modulation, the phenomenon of two-photon process dominates which amplifies photons in the cavity whereas under the weak modulation, the phenomenon of fluorescence dominates. This also shows the balance of energy between different degrees of freedom (optical mode, excitonic mode) of the system. One more thing noticeable here is that the peak intensities of intracavity photons, in general, are more in the good cavity limit.

In the next section, we will study the phenomenon of two-photon process in the same system by considering quantized motion of the movable mirror.                
 
\section{Optomechanical Cavity with Quantized Mirror motion}

In this section, we investigate the same system consisting of QW with an optomechanical oscillator in a light field as shown in figure \ref{b}. In addition, here we consider the quantized mirror motion. Therefore, again due to the rapid motion of the mirror, the cavity frequency sinusoidally modulates with time. In this section, the movable mirror is treated as a quantum mechanical oscillator with frequency $\omega_{m}$. A force proportional to the photon number in the cavity acts on the resonator. This further modulates the various couplings between the different modes of the system. The Hamiltonian with an optomechanical resonator under rotating-wave and dipole approximation is given as
 
\begin{eqnarray}\label{six}\nonumber
H_{II} &=& \hbar\omega_{b}b^{\dagger}b+\hbar\omega_{c}a^{\dagger}a+\frac{1}{2}\hbar\omega_{m}\left(p^{2}+q^{2}\right)+\hbar g_{m}\left(t\right)a^{\dagger}aq+\hbar g_{0}\left(a^{\dagger}b+b^{\dagger}a\right)\nonumber \\ &+& \hbar g\left(t\right)\left(a^{\dagger}b+b^{\dagger}a\right)q+i\hbar\chi\left(t\right) \left(a^{\dagger 2}e^{-2i\omega_{p}t}-a^{2}e^{2i\omega_{p}t}\right)q+i\hbar\epsilon_{p}\left(a^{\dagger}e^{-i\omega_{p}t}-a e^{i\omega_{p}t}\right),
\end{eqnarray}

The derivation of the above Hamiltonian is given in Appendix. Here the third term represents the free energy of the mechanical oscillator. The dimensionless position and momentum operator of the movable mirror are represented by $q$ and $p$ respectively which obeys the commutation relation $\left(\left[q,p\right]=i\hbar\right)$. The fourth term gives the time-dependent interaction energy between the oscillator and cavity photons, where $g_{m}(t)=\omega_{c}\epsilon\sin\left(\Omega t\right)$ is the time-dependent coupling parameter. The sixth term depicts the three-body time-dependent interaction between exciton, cavity photon and mechanical mode with $g(t)=g_{0}\epsilon\sin\left(\Omega t\right)/2$ as time-dependent interaction parameter. A viscious force acts on the mechanical oscillator which damps the mechanical mode with damping rate $\gamma_{m}$. A Brownian stochastic force also effects the mechanical mode with zero-mean $\zeta$. This force is non-Markovian Gaussian noise \citep{abd, gio}. Taking into account all the dissipation processes, the QLE of the system with an optomechanical resonator in the rotating frame at frequency $\omega_{p}$ are given as:

\begin{equation}\label{sevena}
\dot{b}=-i\delta_{b}b-ig_{0}a-ig(t)aq-\gamma b+\sqrt{2\gamma}b_{in},
\end{equation}   

\begin{equation}\label{sevenb}
\dot{a}=-i\Delta a-ig_{m}(t)aq-ig_{0}b-ig(t)bq+2\chi(t)a^{\dagger}q+\epsilon_{p}-\kappa a+\sqrt{2\kappa}a_{in},
\end{equation} 

\begin{equation}\label{sevenc}
\dot{q}=\omega_{m}p,
\end{equation}

\begin{equation}\label{sevend}
\dot{p}=-\omega_{m}q-g_{m}(t)a^{\dagger}a-ig(t)\left(a^{\dagger}b+b^{\dagger}a\right)-i\chi(t)\left(a^{\dagger 2}-a^{2}\right)-\gamma_{m}p+\zeta(t), 
\end{equation}

To a good approximation, the correlation function of the Brownian noise in the limit of a large mechanical quality factor i.e. $\omega_{m}/\gamma_{m} \gg 1$ can be given as \citep{eyo}

\begin{equation}\label{eight}
\left\langle\zeta(t)\zeta(t')\right\rangle\cong\gamma_{m}\left(2n_{th}+1\right) \delta\left(t-t'\right) 
\end{equation}

where $n_{th}=\left[\exp\left(\hbar\omega_{m}/k_{B}T\right)-1\right]^{-1}$ gives the average number of thermal photons with $k_{B}$ representing the Boltzmann constant and $T$ is the temperature of the bath connected to the movable mirror. By using all the correlations of different noises [Eqs. \ref{four} and \ref{eight}] , the coupled differential equations [Eqs. \ref{sevena}-\ref{sevend}] are solved with the help of MATHEMATICA 9.0 to study the effect of time modulation in the system. Here also, we investigate this system under the same regimes as done in the previous section. Apart from the features noticed in the previous section for the photon number inside the cavity and the fluorescent spectrum of light, here we note some additional observations in the plots discussed below.

\begin{figure}[h]
\hspace{-0.0cm}
\begin{tabular}{cc}
\includegraphics [scale=0.75]{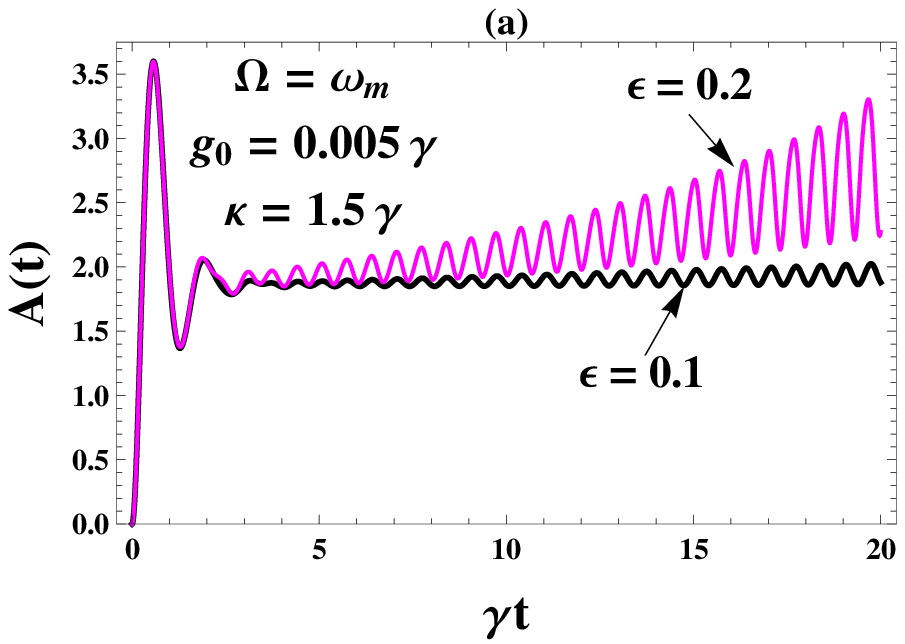}& \includegraphics [scale=0.75] {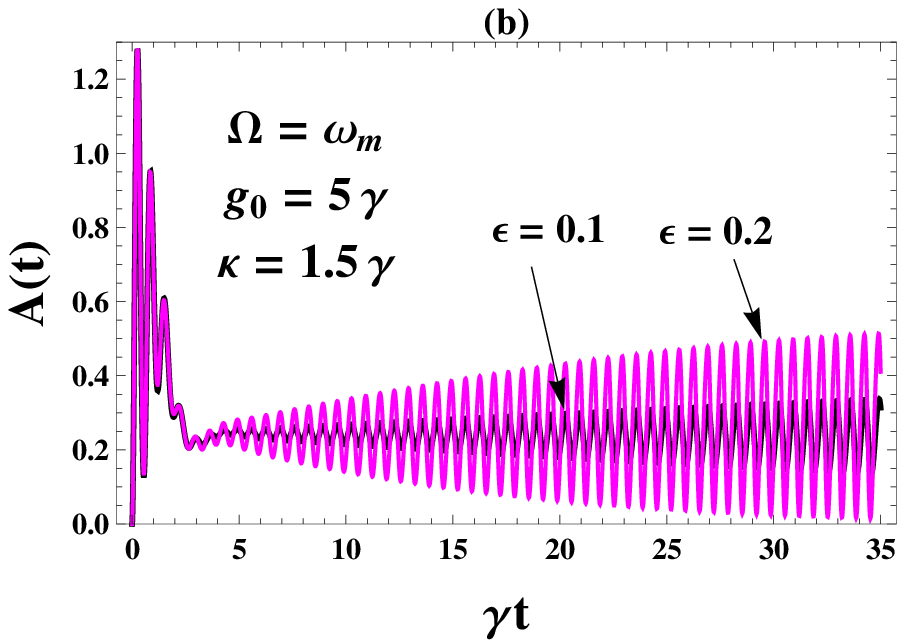}\\
\includegraphics [scale=0.75]{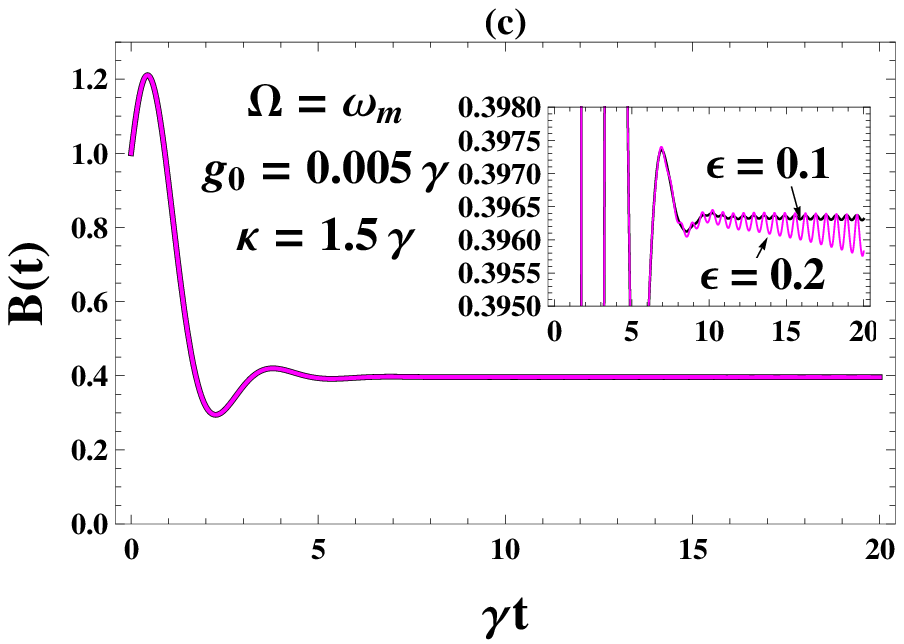}& \includegraphics [scale=0.75] {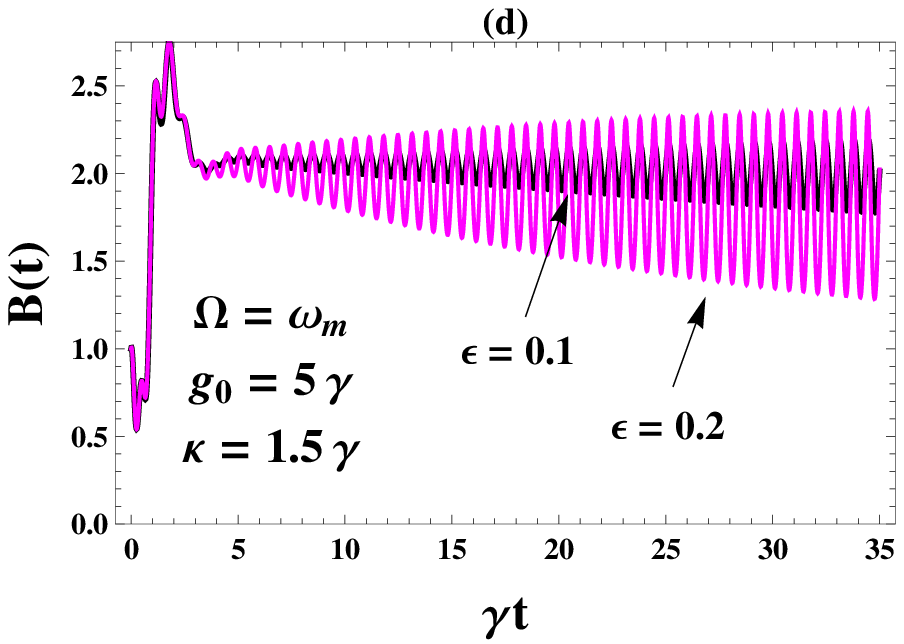}\\
 \end{tabular}
\caption{(Color online) Plots (a) and (b) show the intracavity mean number of photons $\left(A(t)=\langle a^{\dagger}a\rangle\right)$ and plots (c) and (d) show the corresponding fluorescent spectrum of light $\left(B(t)=\langle b^{\dagger}b\rangle\right)$ emitted by excitons in the QW  within an optomechanical cavity versus scaled time $\left(\gamma t\right)$ using time modulated cavity frequency for two modulation amplitudes $\epsilon = 0.1$ (thick line) and $\epsilon = 0.2$ (thin line) in the bad cavity limit with $\kappa>\epsilon \left(\kappa = 1.5\gamma\right)$ at resonant modulating frequency $\left(\Omega=\omega_{m}\right)$. The parameters are $\delta_{b} = 2\gamma$, $\Delta = 4.712\gamma$, $\epsilon_{p} = 5\gamma$, $\gamma_{m} = 10^{-5}\gamma$, $n_{th} = 175$, $\omega_{c} =1.36\gamma$ and $\omega_{m} = 4.712\gamma$. Plots (a) and (c) show the variation of $\left(A(t)=\langle a^{\dagger}a\rangle\right)$ and $\left(B(t)=\langle b^{\dagger}b\rangle\right)$ under the strong modulation $\left(g_{0} = 0.005\gamma\right)$ and plots (b) and (d) show the same variation under the weak modulation $\left(g_{0} = 5\gamma\right)$.}
\label{e}
\end{figure}

In plot \ref{e}, we investigate the system consisting of QW confined within an optomechanical cavity with quantized mirror motion. Plots \ref{e}(a) and \ref{e}(b) illustrate the mean number of intracavity photons $\left(A(t)=\langle a^{\dagger}a\rangle\right)$  with scaled time $\left(\gamma t\right)$ under the strong modulation $(g_{0}=0.005\gamma)$ and weak modulation $(g_{0}=5\gamma)$ respectively in the bad cavity limit $(\kappa\gg\gamma)$ and for $\kappa>\epsilon$ $(\kappa=1.5\gamma)$ at resonant modulating frequency $(\Omega=\omega_{m})$ using two different values of modulation amplitude, $\epsilon=0.1$ (thick line) and $\epsilon=0.2$ (thin line). As time progresses, amplification is observed in the intracavity photon number due to the two-photon process. This additional amplification in photon number in the optomechanical cavity arises due to the external force applied by the movable mirror. In this case, there is an additional degree of freedom (mechanical mode) which exchanges energy with different degrees of freedom of the system. Also, the maximum amount of energy gets transferred between these degrees of freedom at resonant modulating frequency. This is because at resonant frequency, the modulation frequency is same as the natural frequency of the mirror. Furthermore, increase in the modulation amplitude enhances the amplification in the photon number inside the cavity. Also, under the weak modulation, the magnitude of amplification in the intracavity photons is less as compared to that in the strong modulation as the effect of two-photon process is less in the weak modulation. Plots \ref{e}(c) and \ref{e}(d) illustrate the corresponding intensity of fluorescent light emitted by excitons in the QW as a function of scaled time $\left(\gamma t\right)$ under the strong modulation $(g_{0}=0.005\gamma)$ and weak modulation $(g_{0}=5\gamma)$ respectively in the bad cavity limit $(\kappa\gg\gamma)$ and for $\kappa>\epsilon$ $(\kappa=1.5\gamma)$ at resonant frequency $(\Omega=\omega_{m})$ for two modulation amplitudes, $\epsilon=0.1$ (thick line) and $\epsilon=0.2$ (thin line). Plot \ref{e}(c) shows a very small amplification in the intensity of light emitted by excitons in the QW for higher modulation amplitude $(\epsilon=0.2)$. Plot \ref{e}(d) shows the significant amplification in the intensity of light emitted by excitons in the QW under the weak modulation. The increase in the modulation amplitude enhances the amplification in the intensity of light with time. 

\begin{figure}[h]
\hspace{-0.0cm}
\begin{tabular}{cc}
\includegraphics [scale=0.75]{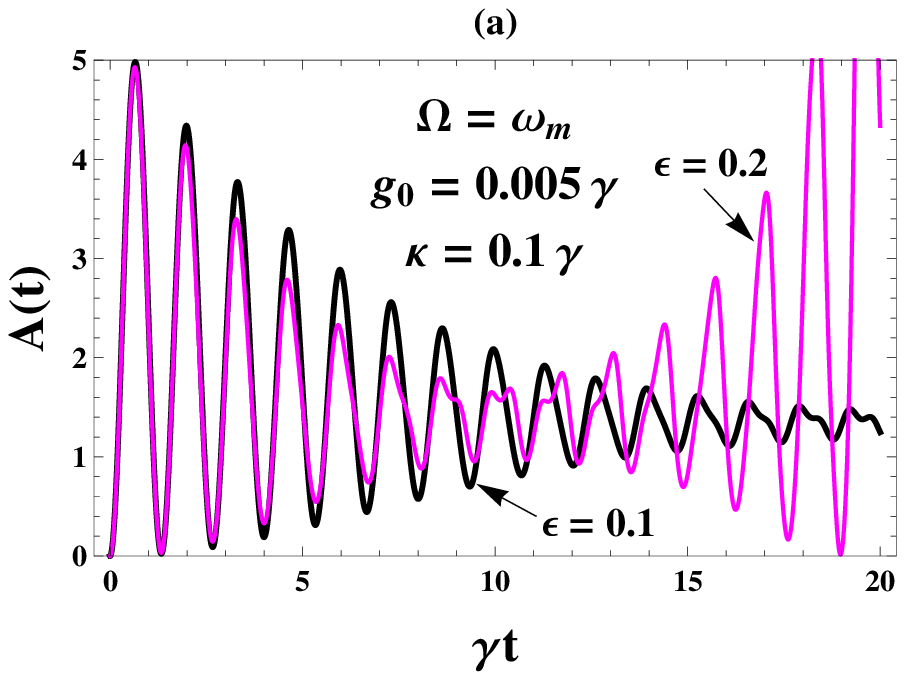}& \includegraphics [scale=0.75] {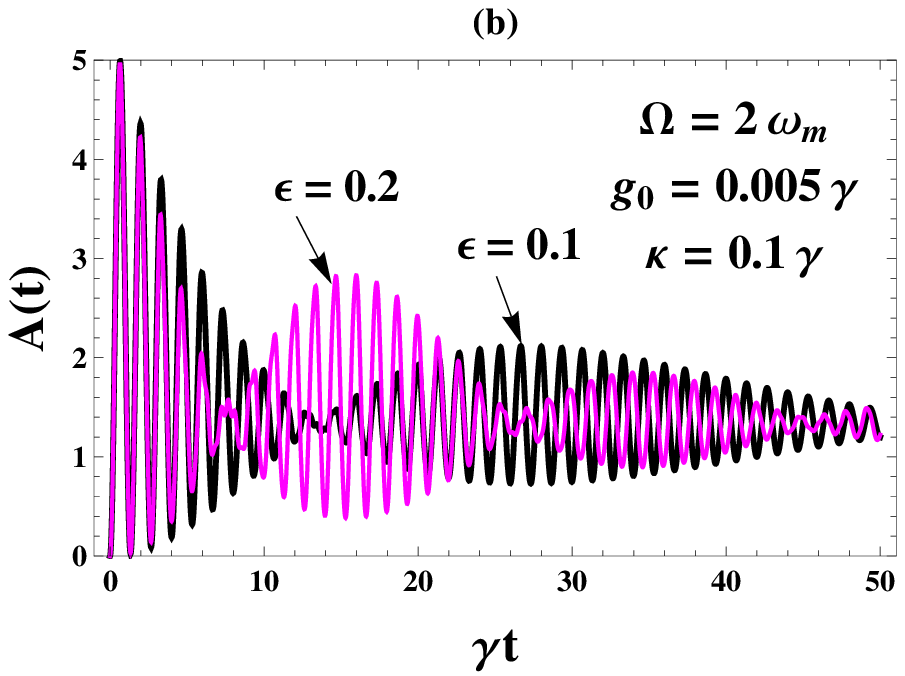}\\
\includegraphics [scale=0.75]{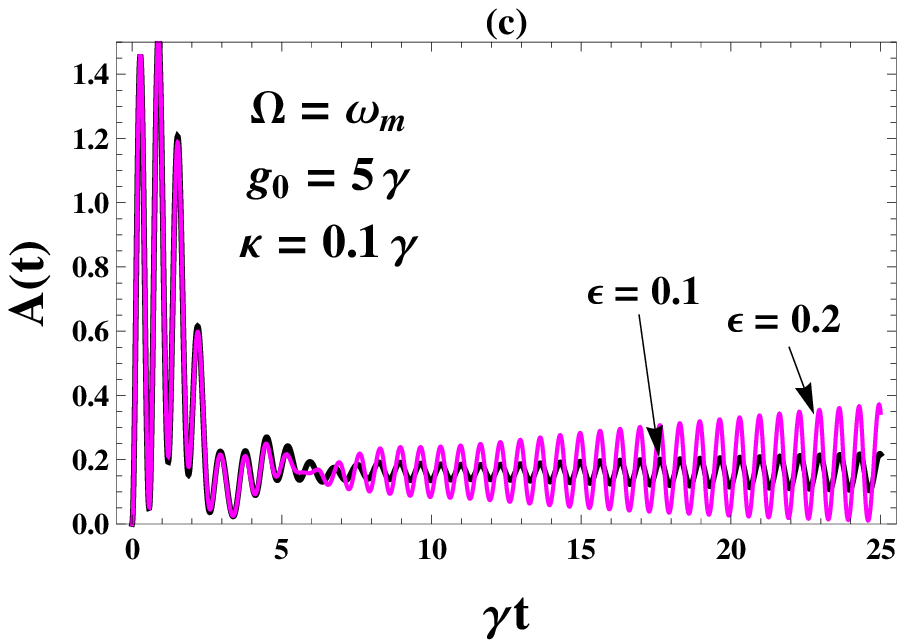}& \includegraphics [scale=0.75] {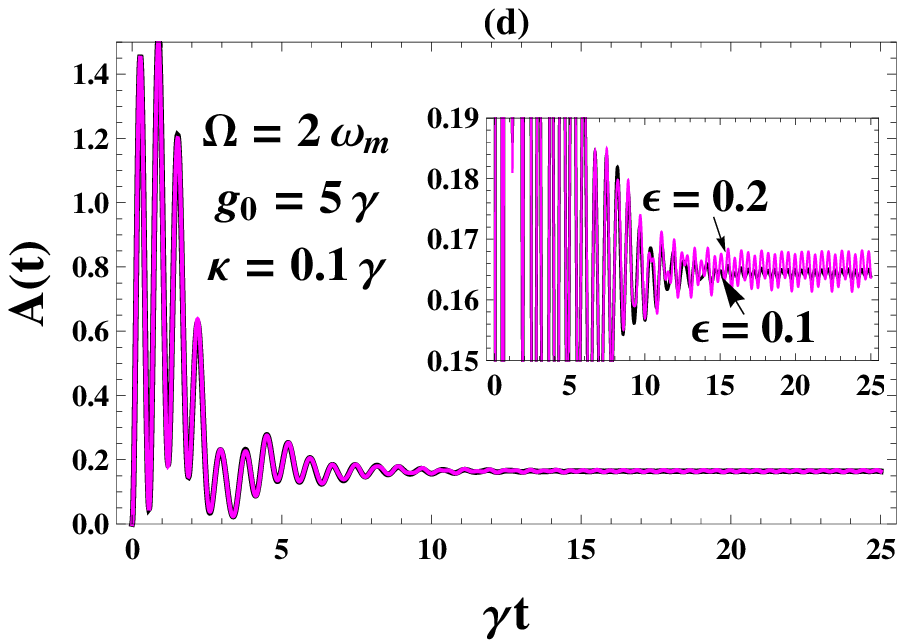}\\
 \end{tabular}
\caption{(Color online) Plots of intracavity mean number of photons $\left(A(t)=\langle a^{\dagger}a\rangle\right)$ inside an optomechanical cavity with scaled time $\left(\gamma t\right)$ using time modulated cavity frequency for two modulation amplitudes $\epsilon = 0.1$ (thick line) and $\epsilon = 0.2$ (thin line) in the good cavity limit with $\kappa\lesssim\epsilon \left(\kappa = 0.1\gamma\right)$. Plots (a) and (b) show the variation of $A(t)$ with time under the strong modulation $\left(g_{0} = 0.005\gamma\right)$ for resonant $\left(\Omega = \omega_{m}\right)$ and off-resonant modulating frequency $\left(\Omega = 2\omega_{m}\right)$ respectively . Plots (c) and (d) show the variation of $A(t)$ with time under the weak modulation $\left(g_{0} = 5\gamma\right)$ for resonant $\left(\Omega = \omega_{m}\right)$ and off-resonant modulating frequency $\left(\Omega = 2\omega_{m}\right)$ respectively. The other parameters used are same as in figure \ref{e}.}
\label{f}
\end{figure}

\begin{figure}[h]
\hspace{-0.0cm}
\begin{tabular}{cc}
\includegraphics [scale=0.75]{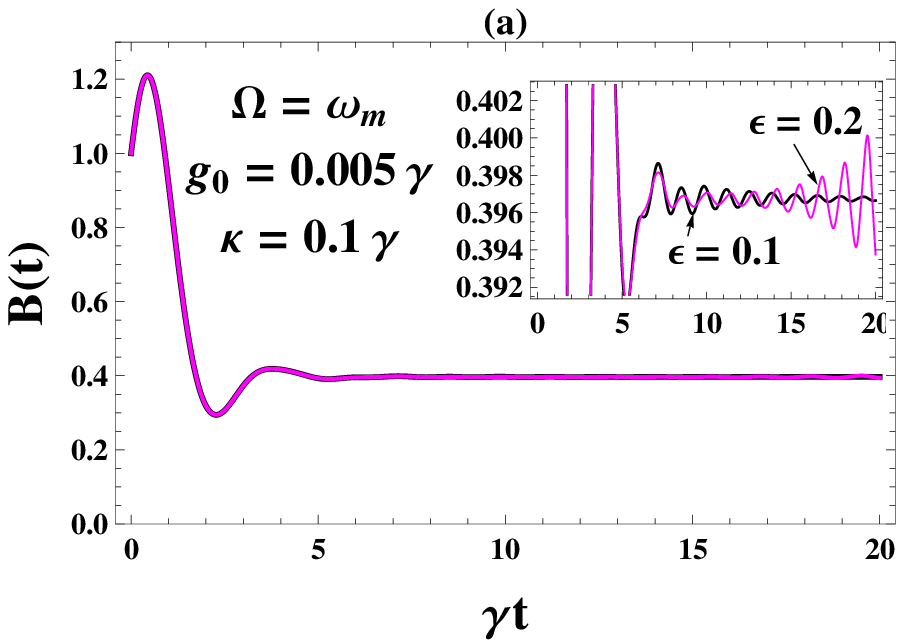}& \includegraphics [scale=0.75] {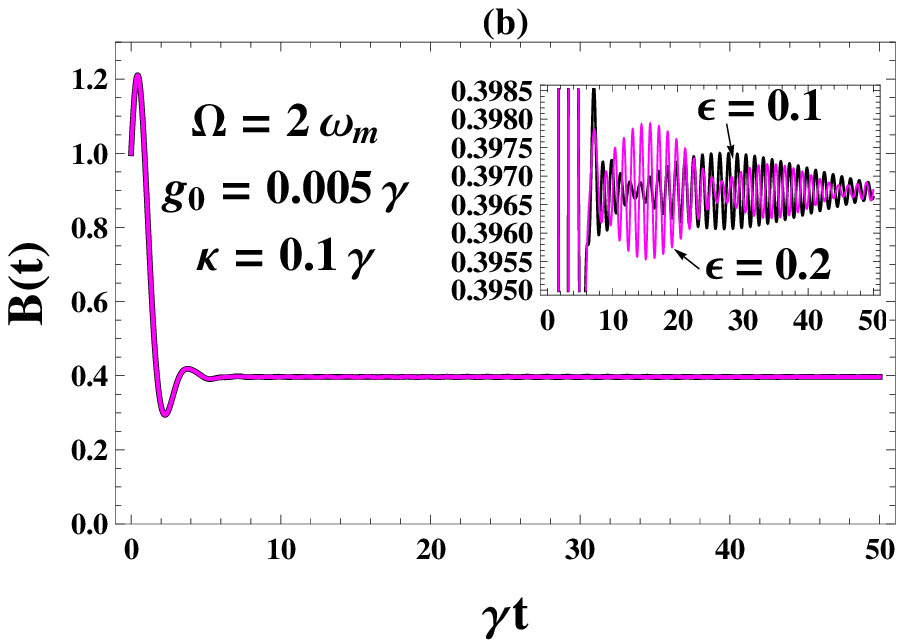}\\
\includegraphics [scale=0.75]{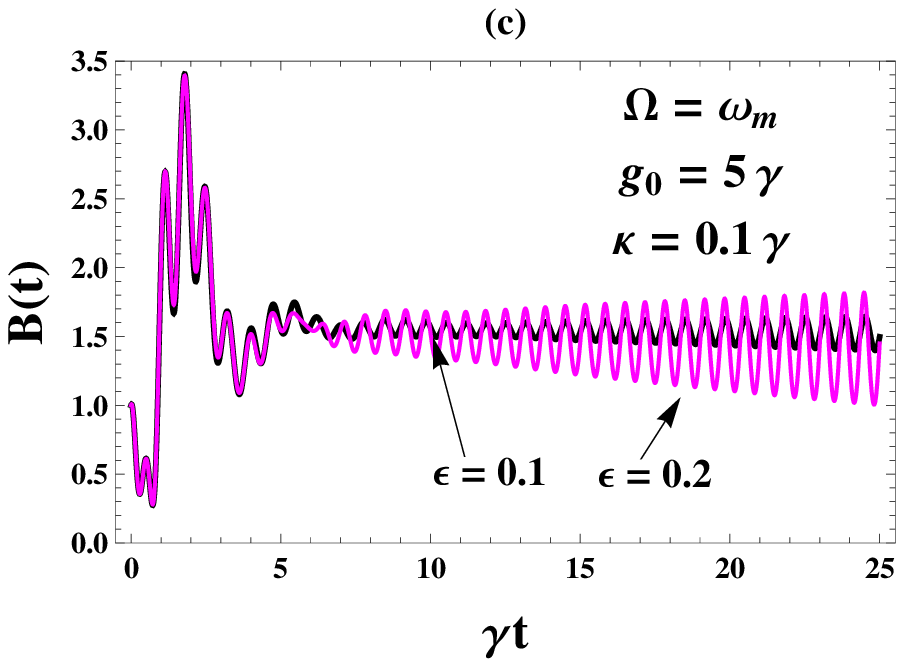}& \includegraphics [scale=0.75] {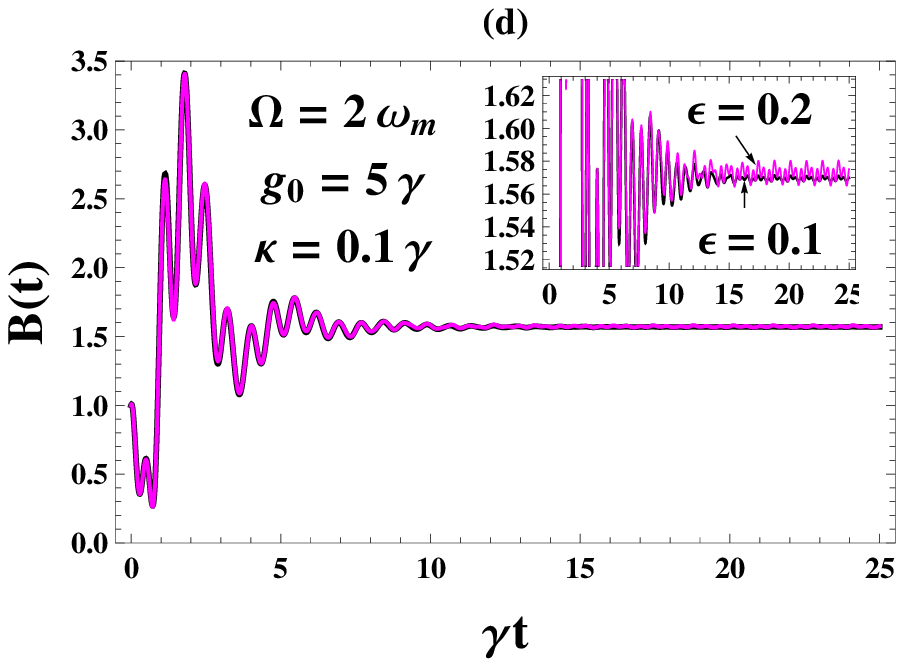}\\
 \end{tabular}
\caption{(Color online) Plots of intensity of fluorescent light $\left(B(t)=\langle b^{\dagger}b\rangle\right)$ vs scaled time $\left(\gamma t\right)$ inside an optomechanical cavity using time modulated cavity frequency for two modulation amplitudes $\epsilon = 0.1$ (thick line) and $\epsilon = 0.2$ (thin line) in the good cavity limit with $\kappa\lesssim\epsilon \left(\kappa = 0.1\gamma\right)$. Plots (a) and (b) show the variation of $B(t)$ with time under the strong modulation $\left(g_{0} = 0.005\gamma\right)$ for resonant $\left(\Omega = \omega_{m}\right)$ and off-resonant modulating frequency $\left(\Omega = 2\omega_{m}\right)$ respectively . Plots (c) and (d) show the variation of $B(t)$ with time under the weak modulation $\left(g_{0} = 5\gamma\right)$ for resonant $\left(\Omega = \omega_{m}\right)$ and off-resonant modulating frequency $\left(\Omega = 2\omega_{m}\right)$ respectively. The other parameters used are same as in figure \ref{e}.}
\label{h}
\end{figure}

Figure \ref{f} shows the mean number of photons inside the optomechanical cavity with scaled time $\left(\gamma t\right)$ under strong and weak modulation in the good cavity limit $(\kappa\ll\gamma)$ and for $\kappa\lesssim\epsilon$ $(\kappa=0.1\gamma)$ using two different values of modulation amplitude, $\epsilon=0.1$ (thick line) and $\epsilon=0.2$ (thin line). Plot \ref{f}(a) illustrates the case of strong modulation $(g_{0}=0.005\gamma)$ at resonant modulating frequency $\left(\Omega=\omega_{m}\right)$. It depicts that as time progresses, the mean number of photons inside the optomechanical cavity firstly decreases and then increases (revival of oscillations). This clearly shows the character of two-photon process as there is amplification in the intracavity photons with time. Fig \ref{f}(b) demonstrates the off-resonant case $\left(\Omega=2\omega_{m}\right)$ under the strong modulation $(g_{0}=0.005\gamma)$. It clearly shows a wavepacket like behaviour with periodic increase and decrease in the intracavity photon number for both the modulation amplitudes. Moreover, here amplification is observed much faster for higher modulation amplitude $(\epsilon=0.2)$. Plots \ref{f}(c) and \ref{f}(d) show the case of weak modulation $(g_{0}=5\gamma)$ at resonant modulating frequency $\left(\Omega=\omega_{m}\right)$ and off-resonant frequency $\left(\Omega=2\omega_{m}\right)$ respectively. At resonant frequency, we are able to distinguish the plots corresponding to two modulation amplitudes (see fig. \ref{f}(c)) while at off-resonant frequency, we are unable to distinguish these plots (see fig. \ref{f}(d)).  

Plot \ref{h} illustrates the intensity of fluorescent light emitted by the excitons in the QW with scaled time $(\gamma t)$ under the strong and weak modulation in the good cavity limit $(\kappa\ll\gamma)$ and for $\kappa\lesssim\epsilon$ $(\kappa=0.1\gamma)$ using two modulation amplitudes, $\epsilon=0.1$ (thick line) and $\epsilon=0.2$ (thin line). Figures \ref{h}(a) and \ref{h}(b) show the case of strong modulation $(g_{0}=0.005\gamma)$ for resonant $\left(\Omega=\omega_{m}\right)$ and off-resonant $\left(\Omega = 2\omega_{m}\right)$ frequency respectively. Figures \ref{h}(c) and \ref{h}(d) show the case of weak modulation $(g_{0}=5\gamma)$ for resonant $\left(\Omega=\omega_{m}\right)$ and off-resonant $\left(\Omega = 2\omega_{m}\right)$ frequency respectively. The features observed in this case, at resonant modulating frequency (see figs. \ref{h}(a) and \ref{h}(c)), are very similar to the one observed in the bad cavity limit (see figs. \ref{e}(c) and \ref{e}(d)). At off resonant frequency, here we observe wavepacket like behaviour with very small amplitude under the strong modulation (see fig. \ref{h}(b)). Also, least disparity is observed in the oscillations of the intensity of fluorescent light for the two values of modulation amplitudes under the weak modulation (see figure \ref{h}(d)). 

In addition to all the above observations seen in the previous section where we considered classical motion of the mirror, here, we observe some more interesting features using the quantized motion of cavity mirror. Firstly, the amplification is clearly observed in intracavity photons in all the cases. This is due to the fact that the system now possesses an extra mechanical mode arising from the quantized mirror motion. So in this case, the coherent energy exchange takes place among three degrees of freedom. Moreover, the off-resonant case of intracavity photon number (see fig. \ref{f}(b)) in the good cavity limit also shows the amplification and wavepacket like behaviour under the strong modulation which is not observed in the weak modulation. This shows that to observe the amplification in the intracavity photons and the phenomenon of two-photon process effectively in a system composed of QW confined in optomechanical cavity, the motion of the cavity movable mirror should be treated quantum mechanically rather than classically.   

In the next section, we investigate the phonon amplification for the system consisting of QW in an optomechanical cavity with constant cavity frequency and amplitude modulated external pump laser beam.    

\section{Phonon Amplification}

In this section, we study the effect of periodic modulation of external laser pump on different modes of the system consisting of QW in an optomechanical cavity as shown in figure \ref{b}. Here, we consider fixed cavity frequency. Rewriting the Hamiltonian for this case as \citep{eyo,far, mar1, mar2, law2}
 
\begin{eqnarray}\label{nine}
H_{III}=\hbar\omega_{b}b^{\dagger}b+\hbar\omega_{c}a^{\dagger}a+\frac{1}{2}\hbar\omega_{m}\left(p^{2}+q^{2}\right)+\hbar g_{m}a^{\dagger}aq+\hbar g_{0}\left(a^{\dagger}b+b^{\dagger}a\right)+i\hbar\epsilon_{p}'(t)\left(a^{\dagger}e^{-i\omega_{p}t}-a e^{i\omega_{p}t}\right), 
\end{eqnarray}

where \begin{eqnarray}
\epsilon_{p}'(t)=\epsilon_{p}\left(1+\eta\cos\left(\lambda t\right)\right).
\end{eqnarray}

Here $\eta$ represents the ampitude of modulation and $\lambda$ represents the frequency of modulation. In this case, the external driving field undergoes the oscillatory motion but the internal parameters of the system remains unchanged. Now, the QLE for the system in the frame rotating at frequency $\omega_{p}$ are given as follows:

\begin{equation}\label{tena}
\dot{b}=-i\delta_{b}b-ig_{0}a-\gamma b+\sqrt{2\gamma}b_{in},
\end{equation}   

\begin{equation}\label{tenb}
\dot{a}=-i\Delta a-ig_{m}aq-ig_{0}b+\epsilon_{p}\left(1+\eta\cos\left(\lambda t\right)\right)-\kappa a+\sqrt{2\kappa}a_{in},
\end{equation} 

\begin{equation}\label{tenc}
\dot{q}=\omega_{m}p,
\end{equation}

\begin{equation}\label{tend}
\dot{p}=-\omega_{m}q-g_{m}a^{\dagger}a-\gamma_{m}p+\zeta(t), 
\end{equation}

By using the correlations given in Eqns. $\left(\ref{four}\right) $ and $\left( \ref{eight}\right)$, we have solved these coupled differential equations using MATHEMATICA 9.0. We have investigated the position dynamics of the mirror, the intracavity photon number and fluorescent spectrum of light with time for the present system. We have considered the strong and weak modulation of the amplitude modulated laser beam in both the limits of good cavity $(\kappa\ll\omega_{m})$ and bad cavity $(\kappa\gg\omega_{m})$.  

\begin{figure}[h]
\hspace{-0.0cm}
\begin{tabular}{cc}
\includegraphics [scale=0.75]{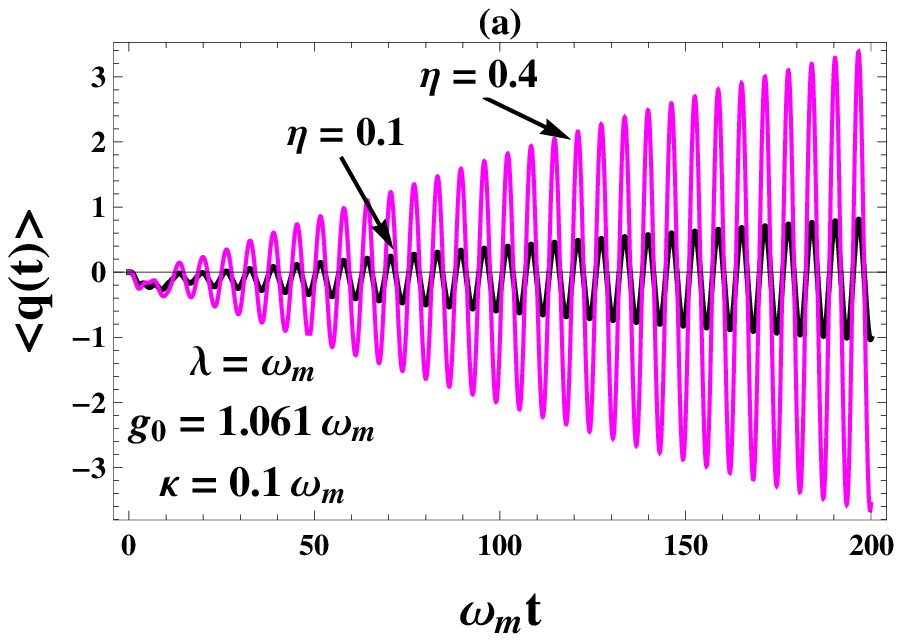}& \includegraphics [scale=0.75] {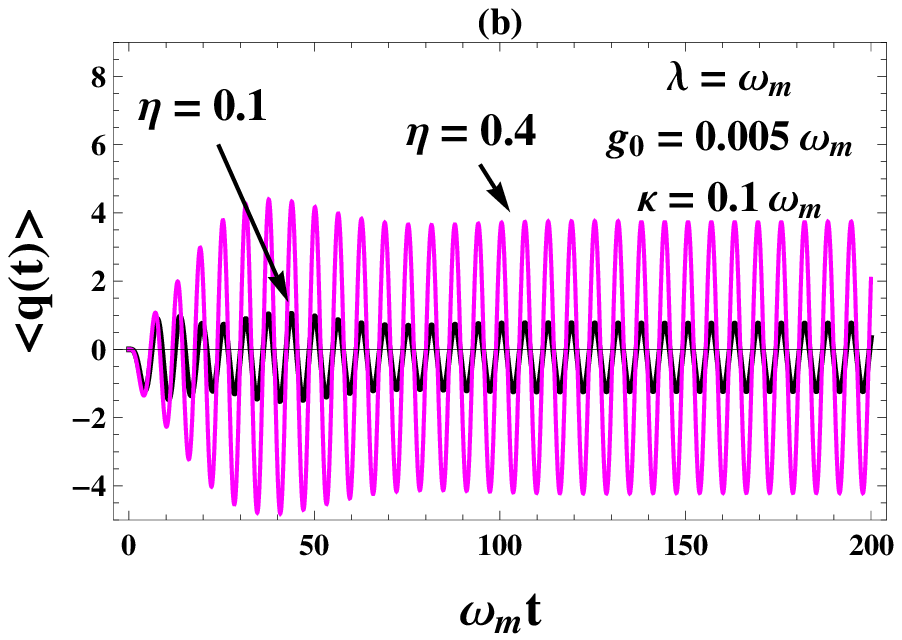}\\
\includegraphics [scale=0.75]{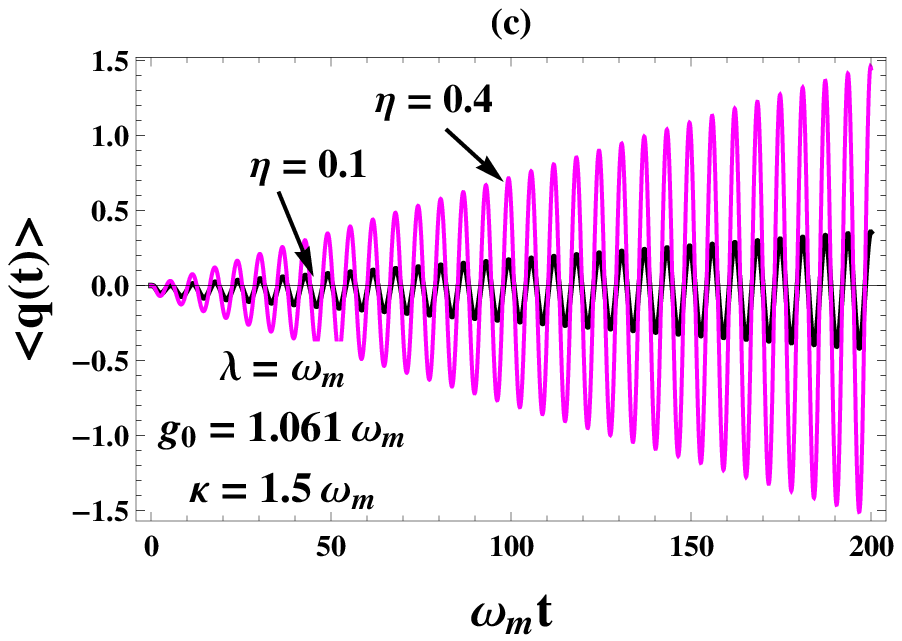}& \includegraphics [scale=0.75] {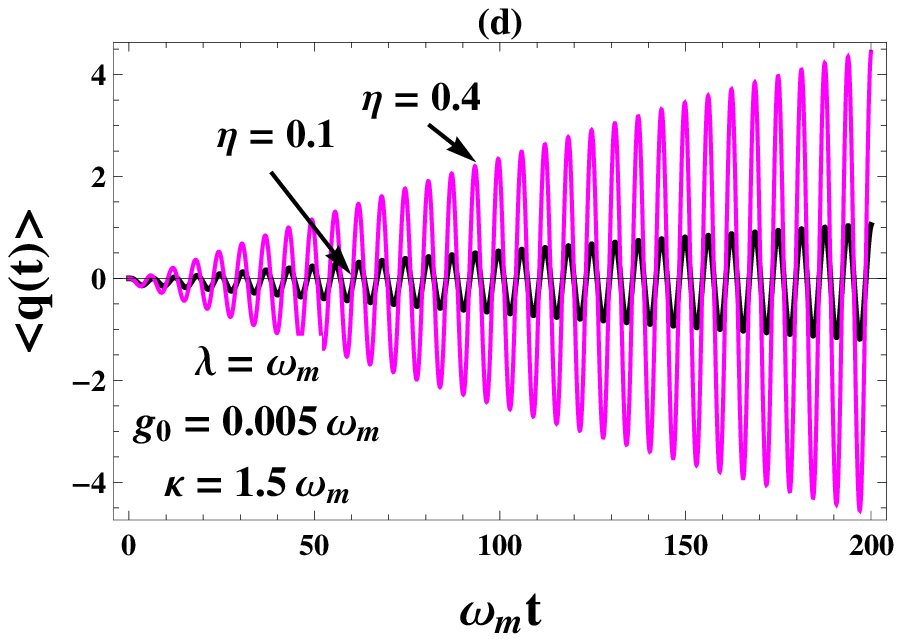}\\
 \end{tabular}
\caption{(Color online) Plots of mean position of mirror $\left\langle q(t)\right\rangle$ of an optomechanical cavity with scaled time $\left(\omega_{m} t\right)$ at resonant modulating frequency $\left(\lambda = \omega_{m}\right)$ for two different modulation amplitudes $\eta = 0.1$ (thick line) and $\eta=0.4$ (thin line) using an amplitude modulated external laser beam. Plots (a) and (b) show the variation of $\left\langle q(t)\right\rangle$ with time for weak modulation $\left(g_{0} = 1.061\omega_{m}\right)$ and strong modulation $\left(g_{0} = 0.005\omega_{m}\right)$ respectively in the good cavity limit with $\kappa\lesssim\epsilon \left(\kappa = 0.1\omega_{m}\right)$. Plots (c) and (d) show the variation of $\left\langle q(t)\right\rangle$ with time for weak modulation $\left(g_{0} = 1.061\omega_{m}\right)$ and strong modulation $\left(g_{0} = 0.005\omega_{m}\right)$ respectively in the bad cavity limit with $\kappa>\epsilon \left(\kappa = 1.5\omega_{m}\right)$. The various parameters used are $\delta_{b} = 0.459\omega_{m}$, $\Delta = \omega_{m}$, $\epsilon_{p} = 1.5\omega_{m}$, $\gamma_{m} = 10^{-5}\omega_{m}$, $n_{th} = 175$, $\gamma =0.212\omega_{m}$ and $g_{m} = 0.1\omega_{m}$.}
\label{i}
\end{figure}

Figure \ref{i} shows the position dynamics $\langle q(t)\rangle$ of the movable mirror with scaled time $\left(\omega_{m}t\right)$ at resonant modulating frequency $\left(\lambda=\omega_{m}\right)$ under strong and weak modulation for two different modulation amplitudes, $\eta=0.1$ (thick line) and $\eta=0.4$ (thin line). Oscillatory behaviour is observed in all the cases. Figures \ref{i}(a) and \ref{i}(b) represent the case of strong and weak modulations respectively in the good cavity limit $(\kappa\ll\omega_{m})$ and for $\kappa\lesssim\epsilon$ $(\kappa=0.1\omega_{m})$. The plots clearly show the amplification in the number of phonons with time. As the modulation amplitude increases the number of phonons in the cavity increases. Plot \ref{i}(a) shows huge enhancement in the phonons with time. Plot \ref{i}(b) demonstrates a slight increase in the phonon number initially. As time passes, the oscillations become steady. Plots \ref{i}(c) and \ref{i}(d) depict the position dynamics for strong and weak modulations respectively in the bad cavity limit $(\kappa\gg\omega_{m})$ and for $\kappa>\epsilon$ $(\kappa=1.5\gamma)$. As observed in the previous case, the phonon number increases with increase in modulation amplitude. In this case, large amplification is observed in the phonons inside the cavity for both the strong and weak modulations. However, the magnitude of amplification in the phonons is more in strong modulation than in the weak modulation.          

\begin{figure}[h]
\hspace{-0.0cm}
\begin{tabular}{cc}
\includegraphics [scale=0.75]{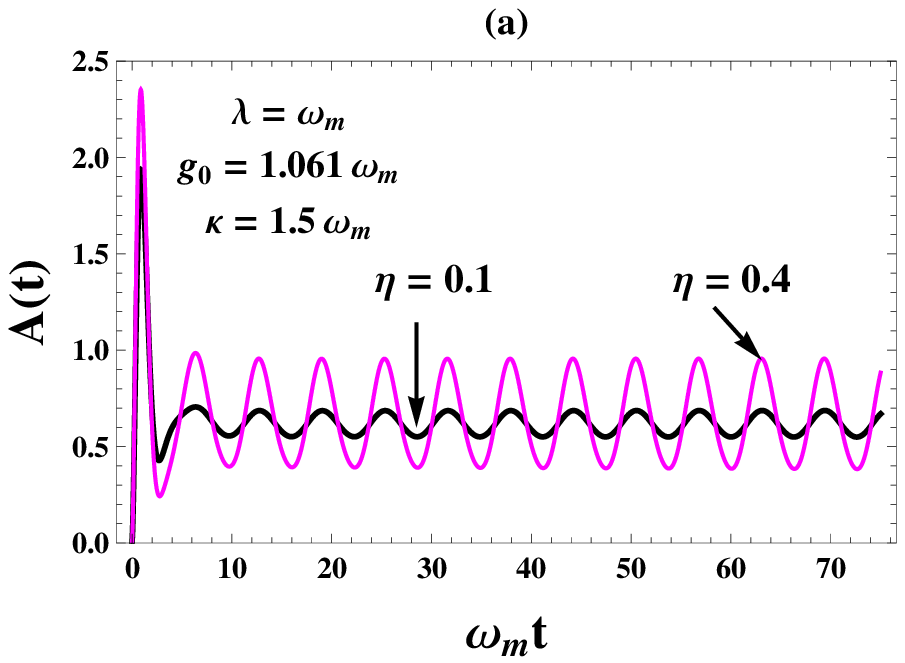}& \includegraphics [scale=0.75] {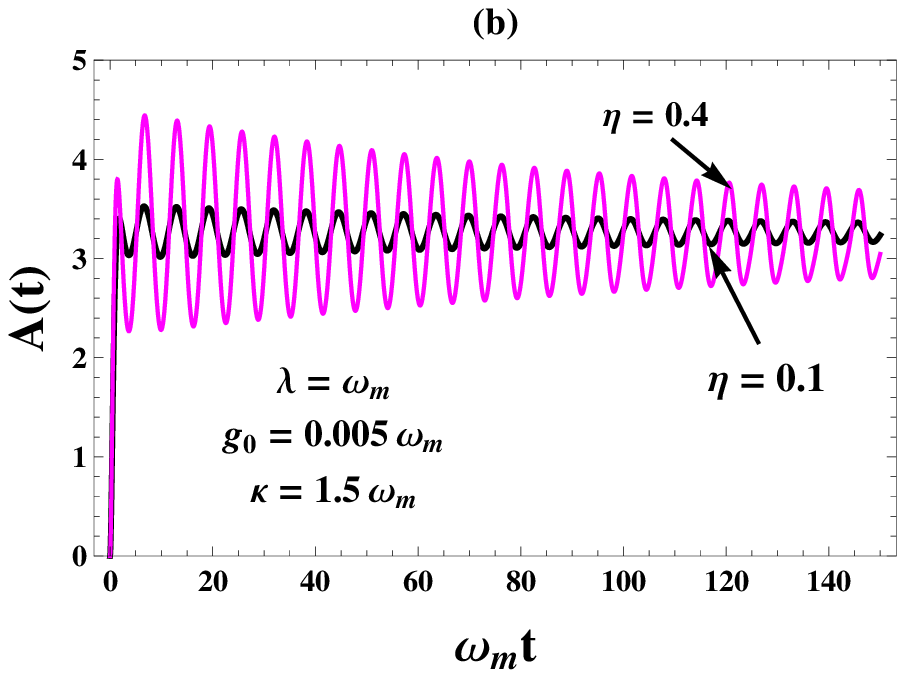}\\
\includegraphics [scale=0.75]{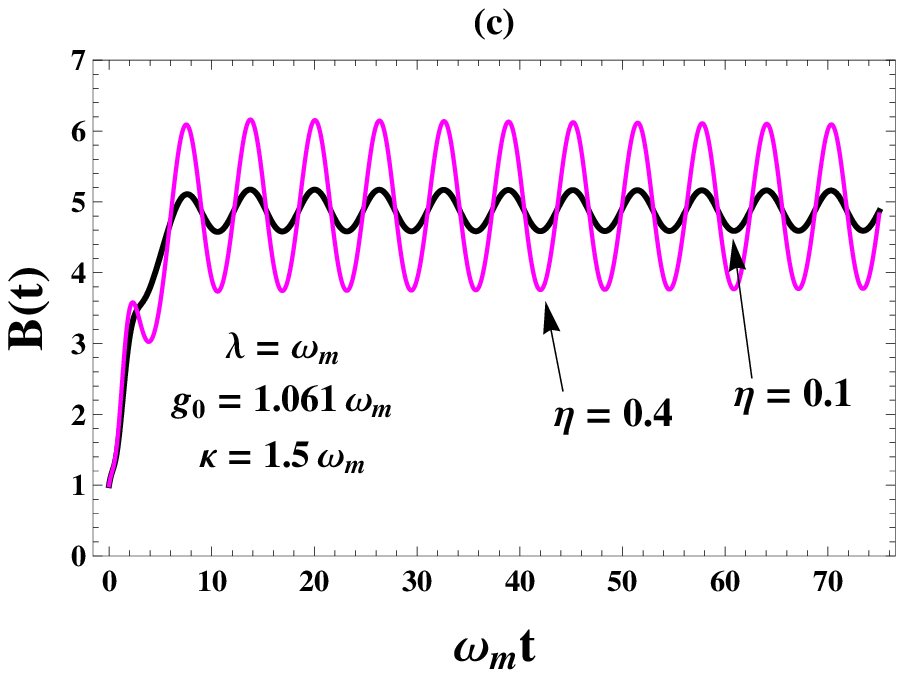}& \includegraphics [scale=0.75] {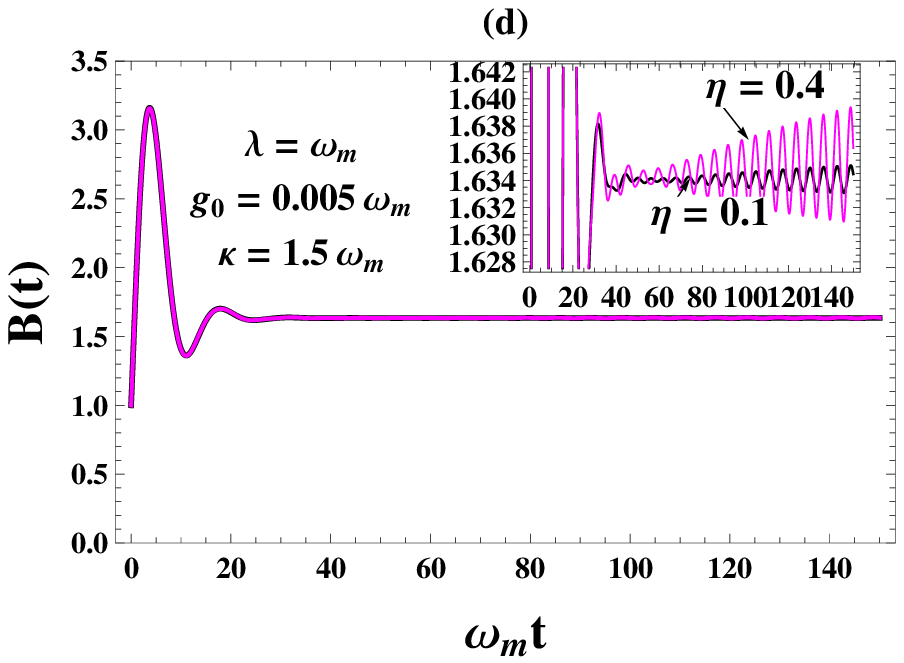}\\
 \end{tabular}
\caption{(Color online) Plots (a) and (b) show the intracavity mean number of photons $\left(A(t)=\langle a^{\dagger}a\rangle\right)$ and plots (c) and (d) show the corresponding fluorescent spectrum of light $\left( B(t)=\langle b^{\dagger}b\rangle\right)$ emitted by excitons in the QW within an optomechanical cavity versus scaled time $\left(\omega_{m} t\right)$ for two modulation amplitudes $\eta = 0.1$ (thick line) and $\eta = 0.4$ (thin line) at resonant modulating frequency $\left(\lambda = \omega_{m}\right)$ in the bad cavity limit with $\kappa>\epsilon \left(\kappa = 1.5\omega_{m}\right)$ using an amplitude modulated external laser beam. The other parameters used are same as in figure \ref{i}.}
\label{j}
\end{figure}

Figures \ref{j}(a) and \ref{j}(b) show the intracavity photon number $(A(t))$ with scaled time $\left(\omega_{m}t\right)$ using amplitude modulated external laser pump beam at resonant modulating frequency $\lambda=\omega_{m}$ in the bad cavity limit $(\kappa\gg\omega_{m})$ and for $\kappa>\epsilon\left(\kappa = 1.5\omega_{m}\right)$ using two modulation amplitudes, $\eta=0.1$ (thick line) and $\eta=0.4$ (thin line). Figure \ref{j}(a) shows the case of weak modulation $\left(g_{0}=1.061\omega_{m}\right)$. It shows a rise in the photon number initially which eventually decreases with time. Periodic oscillations are observed gradually in the mean number of intracavity photons. Again, the amplitude of intracavity photon number increases with higher modulation amplitude. Figure \ref{j}(b) represents the case of strong modulation $\left(g_{0}=0.005\omega_{m}\right)$. Initial rise is observed in the photons inside the cavity which gradually decreases with time. This clearly illustrates the reverse of two-photon process. Plots \ref{j}(c) and \ref{j}(d) show the corresponding fluorescent spectrum of light $(B(t))$ with scaled time $\left(\omega_{m}t\right)$ at resonant modulating frequency $\lambda=\omega_{m}$ in the bad cavity limit $(\kappa\gg\omega_{m})$ and for $\kappa>\epsilon\left(\kappa = 1.5\omega_{m}\right)$ using two modulation amplitudes, $\eta=0.1$ (thick line) and $\eta=0.4$ (thin line). Figure \ref{j}(c) depicts the case of weak modulation $\left(g_{0}=1.061\omega_{m}\right)$. It shows periodic oscillations with an initial rise in the intensity of fluorescent light. This intensity increases with increase in the modulation amplitude. Figure \ref{j}(d) represents the case of strong modulation $\left(g_{0}=0.005\omega_{m}\right)$. This plot demonstrates that the fluorescent spectrum increases initially but gradually becomes steady with time. However, one can notice that there is slight amplification in the intensity of fluorescent light which increases with increase in modulation amplitude. 

Comparing figures \ref{i}(d) and \ref{j}(b), one can clearly notice that there is enhancement in position dynamics of movable mirror and decrease in the intracavity photon number. This shows that the external pump force helps in amplifying phonons. Moreover, the intensity of fluorescent light (see fig. \ref{j}(d)) also shows slight amplification implying that the intracavity photons produced due to two-photon process are converted into fluorescence photons. This clearly illustrates the reverse of two-photon process. 

In our calculations, the experimentally realizable parameters used for all the three systems are discussed as follows. The cavity field can have decay rate $\kappa = 2\pi \times 8.75$ kHz \citep{nag} ($2\pi \times 0.66$ MHz \citep{mur}). The frequency of the mechanical mode in an optomechanical system can be varied from $2\pi \times 100$ Hz \citep{gar}, $2\pi \times 10$ kHz \citep{hun}, to $2\pi \times 73.5$ MHz \citep{schl}. The corresponding damping rate for the optomechanical resonantor can thus be varied from $2\pi \times 10^{-3}$ Hz \citep{gar}, $2\pi \times 3.22$ Hz \citep{hun}, to $2\pi \times 1.3$ kHz \citep{schl}. Here we have also taken an appropriate regime for cooling the optomechanical system to the quantum ground state by considering the cavity-pump detuning to be $\Delta =\omega_{m}$ \citep{schl}. The other parameters used in the present calculations are taken from \citep{eyo, eyo3}.        

\section{Conclusion}
In conclusion, we have proposed a scheme to study the periodic oscillations in a non-stationary system consisting of a QW inside the optomechanical cavity. We have analysed the system with classical cavity mirror motion and with quantized cavity mirror motion. Also, the cavity frequency is periodically modulated with time. The photon number inside the cavities is not steady but it changes with periodic increase and decrease. Amplification is observed in the intracavity photon number with quantized mirror motion. But the photon number inside the cavity with classical mirror motion show periodic oscillations. The fluorescent light emitted by the excitons in the QW has also been investigated. The QW in the optomechanical cavity produces non-periodic damped fluorescent intensity of light when the mirror motion is considered to be classical. However, the fluorescent spectrum of light emitted by excitons in the QW in an optomechanical cavity with quantized mirror motion, shows amplification under certain regimes. It is also observed that under the strong modulation, the phenomenon of two-photon process dominates whereas under the weak modulation, the phenomenon of fluorescence dominates. The system also shows a balance of energy between different degrees of freedom. The present system provides a good way to detect two-photon process at initial stage of fluorescence spectrum. Furthermore, we have also observed reverse of two-photon process for a non-stationary system composed of a QW confined in an optomechanical cavity with an amplitude modulated external laser pump and constant cavity frequency. Significant amplification is observed in the position dynamics of the mirror. The intracavity photons decreases with time. Thus, the modulation in the external laser beam helps in amplifying the phonons. The present scheme can be used as an optical switch with two-photon process as an external control parameter.

\section{Acknowledgements}

Sonam Mahajan acknowledges University of Delhi for the University Teaching Assistantship. A. Bhattacherjee and Neha Aggarwal acknowledge financial support from the Department of Science and Technology, New Delhi for financial assistance vide grant SR/S2/LOP-0034/2010. 

\section{Appendix}

In this Appendix, we derive the time-dependent Hamiltoninan for optomechanical cavity with quantized motion of the movable mirror. As one knows that, the cavity frequency is given as \citep{dod5}

\begin{eqnarray}
\omega_{c}=\frac{C_{1}}{L},
\end{eqnarray}

where $C_{1}$ is constant and $L$ is the length of the cavity. Therefore, change in the cavity length modifies the cavity frequency. Hence, the time varying cavity frequency is given as 

\begin{eqnarray}
\omega_{c}(x,t)=\frac{C_{1}}{L-x(t)}.
\end{eqnarray}

Since the perturbation in the cavity length is very small as compared to its original length. Therefore the time varying cavity frequency becomes

\begin{eqnarray}\label{1}
\omega_{c}(x,t)=\omega_{c}\left(1+\frac{x(t)}{L}\right).
\end{eqnarray}

Now one can take the time-modulated perturbation in cavity length as 

\begin{eqnarray}
x(t)=x'\epsilon '\sin\left(\Omega t\right).
\end{eqnarray}

where $x'=(\Delta x)q$, $x'$ is the quantized perturbation in cavity length, $q$ is the dimensionless position operator of movable mirror, $\epsilon '$ is the modulation amplitude and $\Omega$ is the modulation frequency. Using the time-modulated perturbation of cavity length in equation (\ref{1}), we get

\begin{eqnarray}
\omega_{c}(x,t)=\omega_{c}\left(1+\frac{\left(\Delta x\right)q\epsilon '\sin\left(\Omega t \right)}{L}\right).
\end{eqnarray}

Now we take normalized modulation amplitude as $\epsilon=\left(\left(\Delta x\right)\epsilon '\right)/L$, therefore the time varying cavity frequency becomes

\begin{eqnarray}\label{2}
\omega_{c}(t)=\omega_{c}\left(1+q\epsilon\sin\left(\Omega t \right)\right).
\end{eqnarray}

The effective frequency also changes in this case as it is dependent on time varying cavity frequency (see eqn \ref{fivea}). Therefore, modified effective frequency for the optomechanical cavity becomes

\begin{eqnarray}\label{6}
\chi'(t)=\left(\epsilon'\Omega\cos\left(\Omega t\right)\right)/4=\chi(t)q.
\end{eqnarray}

The coupling constant for the photon-exciton coupling is given as \citep{dod5}

\begin{eqnarray}
g_{0}=\frac{C_{2}}{\sqrt{L}},
\end{eqnarray}

where, $C_{2}$ is a constant. Now the time varying coupling parameter becomes

\begin{eqnarray}
g_{0}(x,t)=\frac{C_{2}}{\sqrt{L-x(t)}}.
\end{eqnarray}

Under the small cavity length perturbation, the time varying coupling parameter becomes

\begin{eqnarray}
g_{0}(x,t)=\frac{C_{2}}{\sqrt{L}}\left(1+\frac{x(t)}{2L}\right).
\end{eqnarray}

Again, using the small quantized perturbation of the cavity length, we get

\begin{eqnarray}
g_{0}(x,t)=g_{0}\left(1+\frac{\left(\Delta x\right)q\epsilon '\sin\left(\Omega t \right)}{2L}\right).
\end{eqnarray}

Therefore, the time varying coupling parameter becomes

\begin{eqnarray}\label{3}
g_{0}(t)=g_{0}\left(1+\frac{q\epsilon\sin\left(\Omega t \right)}{2}\right).
\end{eqnarray}

Hence, the Hamitonian for the optomechanical cavity with quantized mirror motion under rotating-wave and dipole approximation can be written as

\begin{eqnarray}\label{4}\nonumber
H_{II} &=& \hbar\omega_{b}b^{\dagger}b+\hbar\omega_{c}(t)a^{\dagger}a+\frac{1}{2}\hbar\omega_{m}\left(p^{2}+q^{2}\right)+\hbar g_{0}(t)\left(a^{\dagger}b+b^{\dagger}a\right)\nonumber \\ &+& i\hbar\chi'\left(t\right)\left(a^{\dagger 2}e^{-2i\omega_{p}t}-a^{2}e^{2i\omega_{p}t}\right)+i\hbar\epsilon_{p}\left(a^{\dagger}e^{-i\omega_{p}t}-a e^{i\omega_{p}t}\right).
\end{eqnarray}

In the above Hamiltonian, the free energy of the movable mirror is represented by $\frac{1}{2}\hbar\omega_{m}\left(p^{2}+q^{2}\right)$ where $\omega_{m}$ is the mechanical frequency and $p$ is the dimensionless momentum operator of the movable mirror. Now substituting the time varying cavity frequency (eqn. \ref{2}), time varying effective frequency (eqn. \ref{6}) and time varying coupling parameter (eqn. \ref{3}) in the above Hamiltonian (eqn. \ref{4}), the Hamiltonian becomes

\begin{eqnarray}\label{5}\nonumber
H_{II} &=& \hbar\omega_{b}b^{\dagger}b+\hbar\omega_{c}a^{\dagger}a+\frac{1}{2}\hbar\omega_{m}\left(p^{2}+q^{2}\right)+\hbar g_{m}\left(t\right)a^{\dagger}aq+\hbar g_{0}\left(a^{\dagger}b+b^{\dagger}a\right)\nonumber \\ &+& \hbar g\left(t\right)\left(a^{\dagger}b+b^{\dagger}a\right)q+i\hbar\chi\left(t\right) \left(a^{\dagger 2}e^{-2i\omega_{p}t}-a^{2}e^{2i\omega_{p}t}\right)q+i\hbar\epsilon_{p}\left(a^{\dagger}e^{-i\omega_{p}t}-a e^{i\omega_{p}t}\right),
\end{eqnarray}

where $g_{m}(t)=\omega_{c}\epsilon\sin\left(\Omega t\right)$ and $g(t)=g_{0}\epsilon\sin\left(\Omega t\right)/2$ are the time-dependent coupling parameters.

\end{document}